\PassOptionsToPackage{table}{xcolor}
\documentclass[12pt, table]{article}
\usepackage{geometry}
\usepackage{authblk}
\usepackage{hyperref}
\usepackage{appendix}
\usepackage{rotating}
\usepackage{tikz}
\usetikzlibrary{trees}
\usepackage{graphicx, wrapfig}
\graphicspath{{C:/Users/Rem/Documents/Documents/PhD/Thesis/Impacts of a Policy Set on Wealth Distribution Dynamics/Images/}}
\usepackage{amssymb}
\usepackage{amsmath}
\usepackage{mathtools}
\usepackage{amsthm}
\usepackage{esvect}
\usepackage{caption}
\usepackage[table]{xcolor}

\newtheorem{definition}{Definition}

\title{\Large \textbf{Economic Policy Taxonomy}\\ \large Framework for categorizing economic policies}
\author{R. Sadykhov, Dr.G. Goodell and Prof.P. Treleaven}
\affil{University College London}
\date{March 2025}

\begin{document}

\begin{titlepage}
\maketitle
\thispagestyle{empty}
\begin{abstract}

This paper proposes a framework for categorizing economic policies in a form of
a tree taxonomy. The purpose of this approach is to construct an exhaustive
and standardized list of actions that a governing authority has access to and
can change to control an economy. This is advantageous from two perspectives:
by having an exhaustive list of tools, it becomes easier to construct
``complete'' models (i.e., models that take in all empirical data and aim to
simulate economic dynamics) of an economy and understand what the assumptions
of these models are; and by knowing all available actions, economic strategies
can be devised that target specific economic performance metrics with an
exhaustive list of policies.

\end{abstract}
\end{titlepage}

\section{Introduction}
A major problem in economic modelling is the lack of standardization of factors that impact the economic dynamics. Often we ecounter economic models that make implicit assumptions instead of listing all, or at least most, of the assumptions explicitly. This bias often stems from the lack of standardization of assumptions or variables of the model, as the designers of a model implicitly ignore variables without exploring the potential contribution these variables have towards the output of the model or the phenomena being modelled. We believe that one of the main types of such variables in economic models is the set of policies introduced by authorities and regulators (e.g., government, central bank) to control or restrict economic activity of participants in an economy.\par
Another issue with economic policies is the scope at which they are defined. For instance, a central bank can change the \emph{Marginal Lending Rate} in its \emph{Discount Window}, which is an atomic action and cannot be broken down into ``smaller'' actions. However, \emph{Inflation Targeting} is also confusingly called a policy, but its premise is to potentially perform multiple actions such as changing or bounding  \emph{Main Refinancing Rate}, changing \emph{Marginal Lending Rate}, changing \emph{Fractional Reserve Requirements}, and etc. From this example we can see that the term policy is loosely defined, and depending on the situtation it can mean an atomic action or a set of atomic actions. This destinction between existing definitions of a policy makes it even harder to integrate ``policies'' into economic models.\par
Historically, many different atomic policies have been proposed and employed by the governing authorities and it would be very complicated to construct an exhaustive taxonomy that covers all atomic policies that have been utilized as the record about diferent policy implementations is scattered across many different sources. Even if an exhaustive taxonomy can be constructed, it does not limit the real-world gevernments, central banks and regulators from creating new policies for controlling an economy. Therefore, a taxonomy should be flexible enough to allow new atomic policies to be added to it.\par
In trying to address the problems listed above, this paper proposes a way that economic policies can be arranged into a taxonomy, and tightens down some loose ends of the taxonomy tree by defining and listing atomic policies that act as individual actions that could be performed.\par
\subsection{Scope}
The scope of this paper is the analysis and categorization of common atomic economic policies that are widely adopted by governments and  monetary authorities around the world. Furthermore, it is important to note that due to the vast number of atomic actions performed by governing bodies throughout history, almost surely this paper won't list an exhaustive set of atomic policies, rather in this paper we propose a starting point for the further standardization that can be extended.\par
We will start by categorizing common financial operations that are performed by governments and monetary authorities in the form of income statements, which we base on the IFRS Conceptual Framework \cite{IFRSConceptualFramework}. The motivation for this  is to list all the interactions that a governmentt or central bank has with an economy, and therefore, we will see what adjustments can these authorities make to these interactions in order to impact the economy. We will then group the policies based on what parts of the ``Income Statement'' of the authority they affect and based on the standard definitions in economics (e.g., subsidization policies will be categorized under \emph{Fiscal Policies}). For exceptions to these rules, we will justify why they were placed in the specific place inside the taxonomy tree.\par
Lastly, this paper is only concerned with economic policies and does not take into consideration political policies, neither it is aimed at categorizing political policies. This paper does have an overlap with political economics, but we will not be covering political economics in more detail than is needed for this taxonomy. We also are not concerned with the effectiveness of given policies and the payoffs between implementing them in practice.\par
\subsection{Methodology}
Before we start our analysis, we would like to discuss the approach that we will be using and the motivation behind it. To perform the policy categorization, we can employ different approaches, specifically we can cluster atomic policies by their position within the income statements of the government and monetary authorities, or we can cluster them by the targets and effects of these policies in the economy. The first approach can be seen as the top-to-bottom approach as it starts by looking where policies are implemented first, whereas the second approach is bottom-to-top as it starts by identifying the targets of the atomic policies. This difference in approach is very important as the resulting taxonomy can be completely different.\par
In this context, it also important to note that the effects of policies are very context-dependent and the same atomic policies in different environments can lead to entirely different outcomes. In addition, exhaustively listing the effects of each atomic policy can be very complicated and overwhelming, in particular in major economies like the US or UK, where there are too many moving parts to the economy to check for the causality link between an atomic policy and its effect.\par
For these reasons, we decided to use the top-to-bottom approach to perfom the atomic policy categorization, as it provides a clear initial structure to the taxonomy. With the future research, it will be possible to keep listing the effects of different atomic policies, and therefore, refactor the economic policy taxonomy to be bottom-to-top.\par
The preferred method for taxonomising the economic policies depends on the context: top-to-bottom approach is intuitive if we are working with the authorities' income statements, while the bottom-to-top approach is useful if we would like to know all possible controls that can be applied to produced a desired effect. We believe both apoproaches are useful, but only top-to-bottom provides a clear starting point for the future research.

\section{Transaction Channels}
\label{sec:TransactionChannels}
Before devising the taxonomy of economic policies, we would like to take a step back ans ask ``What transactions channels do economic authorities that implement policies have?'', where by ``transaction channel'' we mean a revenue or expense item on the income statement of a government or central bank. The reason we state this question is because most of the ways that a government or monetary authority can impact an economy is through directly interacting with it through these transaction channels. There are of course some exceptions to this rule, but by listing the possible transaction channels first, we can get a good look at the broad picture.\par
Therefore, we categorize the revenue and expense items of a government and central bank into their respective ``Income Statements'', and we aim for these to satisfy the IFRS Conceptual Framework \cite{IFRSConceptualFramework} for financial reporting up to some adjustments (e.g., no tax provision). Another important factor is that some items on these statements should be in the balance sheet under accounts payable or receivable, but since these items will potentially be included into the income statement for the next period we can list them as part of the income statement. For instance, \emph{Personal Income Tax} for a financial period is collected during the next financial period, so until it is collected it is an item on accounts receivable in the government's balance sheet, but the \emph{Personal Income Tax} that has been paid from the previous financial period will be now an item on the income statement.\par
With this in mind, we present a decomposition of government's and central banks's income statements, along with the definitions for each item. The income statements consist out of \emph{Operating Income} (includes revenue and expenses), \emph{Non-Operating Income} and \emph{Irregular Items}, which themselves have further subcategorization.\par
\subsection{Government Income Statement}
\label{subsec:GovernmentIncomeStatement}
In this section we categorized the channels through which the government earns revenue or spends its budget. The items listed under the \emph{Operating Income} are the transaction channels that the government uses as part of its operation, if we can make that analogy to the business corporation. Note that we have split the revenue into two major categories: \emph{Tax Revenue} and \emph{Non-Tax Revenue}, which is the common way to categorize the Government's revenue flows.\par
The \emph{Non-Operating Income} contains the transaction channels that the government has access too, but they are not part of the main operations of the government, and despite items like \emph{Foreign Aid} can be very impactful to government's budget balance, we do not consider such transaction channels as being the government's income for its ``basic operation''.\par
Lastly, \emph{Irregular Items} contains cash flows that the government does not receive regularly and cannot rely upon in most circumstances.\par
\subsubsection{Operating Income}
\begin{itemize}
	\item \textbf{Revenue}
	\begin{itemize}
		\item \textbf{Tax Revenue}: Government's revenue collected from taxes.
		\begin{itemize}
			\item \textbf{Income Tax}: Tax imposed on individuals or entities (i.e., taxpayers) in respect of the income or profits earned by them.
			\begin{itemize}
				\item \textbf{Personal Income Tax}: Tax on income of an individual.
				\item \textbf{Capital Gains Tax}: Tax on profits realized on the sale of a non-inventory asset.
				\item \textbf{Dividend Tax}: Tax imposed on dividends paid by a corporation to its shareholders.
				\item \textbf{Corporate Tax}: Tax levied on the income or capital of corporations and other similar legal entities.
				\item \textbf{Excess Profit Tax}: Tax on any profit over a certain amount, that is usually used to prevent coorporations from profiteering from certain activities (e.g., profiteering from war).
			\end{itemize}
			\item \textbf{Property Tax}: Tax imposed on assets and real estate.
			\begin{itemize}
				\item \textbf{Inheritance Tax}: Tax paid by a person who inherits money or property of a person who has died.
				\item \textbf{Estate Tax}: Levy on the estate (money and property) of a person who has died.
				\item \textbf{Gift Tax}: Tax on money or property that one living person or corporate entity gives to another.
				\item \textbf{Wealth Tax}: Tax levied on the total value of personal assets including bank deposits, real estate, assets in insurance and pension plans, ownership of unincorporated businesses, financial securities, and personal trusts.
				\item \textbf{Net Wealth Tax}: Wealth tax net the liabilities such as mortgages and loans.
				\item \textbf{Real Estate Tax}: Tax on the property, domestic or otherwise.
				\item \textbf{Land Value Tax}: Levy on the value of land without regard to buildings, personal property and other improvements upon it.
				\item \textbf{Expatriation Tax}: Tax on individuals who renounce their citizenship or residence.
				\item \textbf{Franchise Tax}: Tax levied on the net worth of a corporation.
				\item \textbf{Carucage}: Tax collected only when the government requires extra revenue and otherwise is not regularly leveid.
			\end{itemize}
			\item \textbf{Sales Tax}: Tax levied on sales of goods and services.
			\begin{itemize}
				\item \textbf{Value Added Tax}: Tax levied on the price of a product or service at each stage of production, distribution, or sale to the end consumer.
				\item \textbf{Manufacturers' Sale Tax}: Tax on sales of tangible personal property by manufacturers and producers.
				\item \textbf{Wholesale Sales Tax}: Tax on sales of wholesale of tangible personal property when in a form packaged and labeled ready for shipment or delivery to final users and consumers.
				\item \textbf{Retail Sales Tax}: Tax on sales of retail of tangible personal property to final consumers and industrial users.
				\item \textbf{Gross Receipts Tax}: Tax levied on all sales of a business.
				\item \textbf{Luxury Tax}: Tax on luxury goods (i.e., products that are not considered essential).
				\item \textbf{Excise Tax}: Tax on manufactured goods that is normally levied at the moment of production for internal consumption in the economy rather than at sale.
				\item \textbf{Use Tax}: Tax applied where a product or service is purchased and then converted for a different use, without having paid tax when it was initially purchased.
				\item \textbf{Securities Turnover Excise Tax}: Tax on a securities trade.
				\item \textbf{Turnover Tax}: Tax on intermediate and capital goods applicable to a production process or stage.
			\end{itemize}
			\item \textbf{Pigouvian Tax}: Tax on any market activity that generates negative externalities (i.e., external costs incurred by the producer that are not included in the market price).
			\begin{itemize}
				\item \textbf{Environmental Tax}: Tax levied on activities which are considered to be harmful to the environment.
				\item \textbf{Congestion Tax}: Tax charged from users of public goods that are subject to congestion through excess demand.
				\item \textbf{Pigouvian Good Tax}: Tax on a good that produces or results in negative externalities (e.g., fat tax, alcohol tax, sugar tax).
				\item \textbf{Pigouvian Service Tax}: Tax on a service that produces or results in negative externalities (e.g., fat tax, alcohol tax, sugar tax).
			\end{itemize}
			\item \textbf{Scutage}: Tax levied to bypass military or civil service.
			\item \textbf{Bank Tax}: Tax levied on the capital at risk of financial institutions, excluding government insured deposits.
			\item \textbf{Tariff}: Tax imposed by the government of a country or by a supranational union on imports or exports of goods.
		\end{itemize}
		\item \textbf{Non-Tax Revenue}: Government's revenue not generated by taxes.
		\begin{itemize}
			\item \textbf{State-Owned Enterprises' Revenue}: Revenue from the services run by the government.
			\begin{itemize}
				\item {User Fees}: Fee paid by a facility user as a necessary condition for using the facility
			\end{itemize}
			\item \textbf{Rents and Royalties from Private Companies}: Fess for using or improving public land by private companies (e.g., extraction of minerals).
			\item \textbf{Fines}: Penalty fees levied from criminal offenders.
			\item \textbf{Permits and Licences Fees}: Fees for granting permits and licences.
			\item \textbf{Intragovernmental Revenue}: Aid from another level of government or from equalization payments.
			\begin{itemize}
				\item \textbf{Intragovernmental Aid}: Aid provided by another level of government structure.
				\item \textbf{Equalization Payments}: Payments made from the federal government to subnational governments with the objective of offsetting differences in available revenue or in the cost of providing services.
			\end{itemize}
			\item \textbf{Seignoirage}: Profit from the difference between the value of money and the cost to produce and distribute it.
			\item \textbf{Financing Revenue}: Government revenue raised through financing operations (i.e., issuance of debt or equity).
			\begin{itemize}
				\item \textbf{International Government Debt Issuance}: Funds borrowed from foreign creditors.
				\item \textbf{Domestic Government Debt Issuance}: Funds borrowed from domestic creditors.
				\item \textbf{Equity}: Sovereign selling the future tax liability discharge credit in exchange for immediate financing.
			\end{itemize}
		\end{itemize}
	\end{itemize}
	\item \textbf{Expenses}
	\begin{itemize}
		\item \textbf{Cost of Sales}
		\begin{itemize}
			\item \textbf{State-Owned Enterprises's Cost of Sales}: Cost of sales assocciated with the services run by the governments.
			\item \textbf{Wage Accrual}: Expenditure on wages of government employed people.
		\end{itemize}
		\item \textbf{Other Expenses}
		\begin{itemize}
			\item \textbf{Government Grant}: Funds issued by the government for a specific purpose linked to public benefit.
			\item \textbf{Subsidization}: Issuance of funds to individuals and households, as well as businesses with the aim of stabilizing the economy.
			\begin{itemize}
				\item \textbf{Production Subsidization}: Issuance of funds to encourages suppliers to increase the output of a particular product by partially offsetting the production costs or losses.
				\item \textbf{Consumption Subsidization}: Issuance of funds directly to consumers to encourage consumption of a specific good or service.
				\item \textbf{Price Subsidization}: Issuance of funds to businesses to pay for part of the cost of the good or service to prevent price inflation for consumers.
				\item \textbf{Employment Subsidization}: Issuance of funds to businesses to maintain the employment level, in particular during recession.
				\item \textbf{Negative Tax Payments}: Payments made by the government to taxpayer if the taxpayer satisfies certain requirements.
				\item \textbf{Import Subsidization}: Issuance of funds to cover part of the cost of the imported goods and services for the domestic consumers.
				\item \textbf{Export Subsidization}: Issuance of funds to support production of goods and services that a country is producing to be exported.
			\end{itemize}
		\end{itemize}
		\item \textbf{Depreciation Expenditure}: Expenditure that covers the cost of depreciation of fixed assets owned by the government.
	\end{itemize}
\end{itemize}
\subsubsection{Non-Operating Income}
\begin{itemize}
	\item \textbf{Sovereign Wealth Fund Profit (Net)}: Net income from a sovereign wealth fund.
	\item \textbf{Foreign Aid}: Voluntary transfer of resources from one country to another.
	\item \textbf{Interest Expense}: Interest Expense on the outstanding sovereign debt.
\end{itemize}
\subsubsection{Irregular Items}
\begin{itemize}
	\item \textbf{Indemnity and War Reparations (Net)}: Net income from war compensations.
	\item \textbf{Sale and Purchase of Fixed Assets}: Sale of government owned fixed assets, or the act of purchase of fixed assets from the private sector by the government.
	\item \textbf{Voluntary Donations}: Optional monetary contributions to the government.
\end{itemize}
\subsection{Monetary Authority Income Statement}
\label{subsec:MonetaryAuthorityIncomeStatement}
Similarly to the previous section, in this section we present the income statement decomposition of another economic policy-maker - central bank. In this paper we will be referring to monetary authority and central bank interchangeably.\par
Here we will list all the common transaction channels used by the central bank, and state their definitions. Note, that we have kept the structure of the income statement with \emph{Operating Income}, \emph{Non-Operating Income}, \emph{Irregular Items} being its subcategories that carry the same meaning as in the previous section.\par
\subsubsection{Operating Income}
\begin{itemize}
	\item \textbf{Revenue}
	\begin{itemize}
		\item \textbf{Sale of Government Securities}: Sale of government securities in the open market.
		\item \textbf{Loans}: Loans issued to the central bank account holders.
		\begin{itemize}
			\item \textbf{Personal Loans}: Loans to individuals.
			\item \textbf{Business Loans}: Loans to businesses.
			\item \textbf{Discount Window Loans}: Loans issued to eligible institutions in exchange for a collateral, usually on a short-term basis, to meet temporary shortages of liquidity caused by internal or external disruptions.
		\end{itemize}
		\item \textbf{Sale of Foreign Currency}
		\begin{itemize}
			\item \textbf{Sale of Foreign Currency Reserve}: Sale of a foreign currency from the central bank reserve for the domestic currency.
			\item \textbf{Foreign Currency Agreements Delivery}: Delivery of foreign currency to the counterparty in exchange for domestic currency (e.g., swaps, forwards, futures).
		\end{itemize}
		\item \textbf{Sale of Commodity Reserve}: Sale of a commodity from the central bank reserve for the domestic currency.
	\end{itemize}
	\item \textbf{Expenses}
	\begin{itemize}
		\item \textbf{Purchase of Government Securities}: Purchase of government securities in the open market.
		\item \textbf{Interest Expense}: Interest Expense on loans taken.
		\item \textbf{Purchase of Foreign Currency}
		\begin{itemize}
			\item \textbf{Purchase of Foreign Currency Reserve}: Purchase of a foreign currency for the domestic currency.
			\item \textbf{Foreign Currency Agreements Receival}: Receival of foreign currency from the counterparty in exchange for domestic currency (e.g., swaps, forwards, futures).
		\end{itemize}
		\item \textbf{Purchase of Commodity Reserve}: Purchase of a commodity for the domestic currency.
		\item \textbf{Depreciation Expenditure}: Expenditure that covers the cost of depreciation of fixed assets owned by the central bank.
	\end{itemize}
\end{itemize}
\subsubsection{Non-Operating Income}
\begin{itemize}
	\item \textbf{Profit from Investing Activities (Net)}: Net income generated by central bank performing open market operations.
\end{itemize}
\subsubsection{Irregular Items}
\begin{itemize}
	\item \textbf{Helicopter Money}: Process of central bank creating money (without assets as a counterpart) in its balance sheet and issuing it directly to households.
\end{itemize}

\section{Taxonomy Root}
\label{sec:TaxonomyRoot}
Having listed the incoming and outgoing cash flows of a government and central bank, we now develop a taxonomy of economic policies on the top of that. Our goal is to propose a robust framework for categorizing economic policies, but we also want it to integrate well with the current economics theory, so we proceed by filtering economic policies into already well-defined parent categories.\par
However, first, we must introduce a definition of policy that we can refer to throughout the paper:
\begin{definition}
\label{def:Policy}
	\textbf{Policy} is a set of ideas or a plan of what to do in particular situations that has been agreed to officially by a group of people, a business organization, a government, or a political party \cite{CambridgeDictionary}.
\end{definition}
This definition defines policy as a set of actions that an official body such as government may perform to achieve their desired goals, but we also must define the individual actions and changes that can be implemented (i.e., atomic policies):
\begin{definition}
\label{def:AtomicPolicy}
	\textbf{Atomic Policy} is an individual action or idea that applies a change to a specific body or phenomena.
\end{definition}
To highlight what we mean by an atomic policy, let us demonstrate it through an example. Assume a government needs to raise extra revenue, so it introduces changes to its \emph{Monetary Policy}, and one of the changes that could be proposed is the increase in \emph{Personal Income Tax} by increasing its tax rate. In this example, \emph{Monetary Policy} is the example of a policy as per definition \ref{def:Policy}, whereas the new tax rate for the \emph{Personal Income Tax} is an example of an atomic policy as per definition \ref{def:AtomicPolicy}. Our objective in the sections that follow is to describe atomic policies and categorize them under the broader policy sets such as \emph{Fiscal Policy}, \emph{Monetary Policy} or \emph{International Trade Policy}.\par
A commonly defined category of an economic policy is the \emph{Stabilization Policy} (i.e., fiscal and monetary policies) that aims to stabilize the transition of an economy through the economic cycle, while also stabilizing inflation and employment. Another common subtype of economic policy is the \emph{International Trade Policy}, which is used by the government to facilitate international trade, access foreign labour and resources, and stabilize domestic production through the regulation of trading activity.\par
The following are the definitions of the \emph{Stabilization Policy} and \emph{International Trade Policy}:
\begin{definition}
\label{def:StabilizationPolicy}
	\textbf{Stabilization Policy} is a package or set of measures introduced to correct normal behaviour of the business or credit cycle in a financial system or economy.
\end{definition}
\begin{definition}
\label{def:InternationalTradePolicy}
	\textbf{International Trade Policy (Trade Policy, Commercial Policy)} is a government's set of rules governing international trade.
\end{definition}
From the definitions above we see that \emph{Stabilization Policy} is directed at domestic economy, whereas the \emph{International Trade Policy} is directed at the balance of exports and imports. The two categories of policies can be described as affecting different components of aggregate demand, defined by equation \ref{eq:AggregateDemand}.
\begin{equation}
\label{eq:AggregateDemand}
\begin{split}
	\textit{Aggregate Demand} &= \textit{Consumption} + \textit{Investment} + \textit{Government Spending} +\\
	&+ (\textit{Export} - \textit{Import})
\end{split}
\end{equation}
Aggregate demand splits an economy into 5 sectors: Consumption (i.e., households expenses), Investment (i.e., business expenses, property sales), Government Spending, Export and Import (note that Investment excludes purchase of financial products to avoid double counting). The \emph{International Trade Policy} affects the Export and Import components of the aggregate demand, while the \emph{Stabilization Policy} is concerned with Consumption, Investment and Government Spending. Thus we use this as a rule to create the first ramification under the \emph{Economic Policy}.\par
We also need to make further sub-categorization of different types of the stabilization policies, which are \emph{Fiscal Policies} and \emph{Monetary Policies}, and they are defined as follows:
\begin{definition}
\label{def:FiscalPolicy}
	\textbf{Fiscal Policy} is the use of government revenue collection and expenditure to influence a country's economy.
\end{definition}
\begin{definition}
\label{def:MonetaryPolicy}
	\textbf{Monetary Policy} is the policy adopted by the monetary authority to affect monetary conditions to accomplish economic objectives such as high employment and price stability.
\end{definition}
Both types of stabilization policies are used to correct the economy's transition through economic cycle while stabilizing inflation and employment, but they differ in their implementation mechanism. \emph{Fiscal Policy} is introduced throught the government expenditure and revenue collection (e.g., taxation) in an attempt to rebalance Consumption and Investment, while \emph{Monetary Policy} operates through making changes to financial conditions in form of controlling lending and borrowing, purchasing and selling of financial assets, and manipulating exchnage rates. Most notably, \emph{Fiscal Policy} is devised by the government, but \emph{Monetary Policy} is devised by the monetary authority of a country, which is usually a central bank.\par
Having defned the initial subcategorization in the taxonomy tree, we will now further examine the sub-branches of \emph{International Trade Policy}, \emph{Fiscal Policy} and \emph{Monetary Policy}. Most of the atomic policies in these  sub-branches can easily be grouped under a parent category (e.g., \emph{Taxation}) and we will not be expanding on why those policies have been grouped together as it is trivial, but for more diverse groups of policies we present the reasoning behiond their categorization.\par

\section{Fiscal Policy}
\label{sec:FiscalPolicy}
From the definition \ref{def:FiscalPolicy} we know that \emph{Fiscal Policy} is about collecting government revenue through different channels and then redistributing it to other parts of an economy in an attempt to correct the economic cycle and stabilize inflation and employment. We note that the atomic policies inside the \emph{Fiscal Policy} primarily manipulate the items defined in section \ref{subsec:GovernmentIncomeStatement} to achieve the economic stability. There are some exceptions to this rule, as there exist atomic policies that do not directly ajust the government's income statement, but act as auxiliary tools to steer an economy (e.g., \emph{Labour Unionization}).\par
Given the structure of government's incoming and outgoing cash flows in section \ref{subsec:GovernmentIncomeStatement}, we propose the following subdivision of \emph{Fiscal Policies}:
\begin{itemize}
	\item \textbf{Revenue Policy}: Policies that directly impact the revenue collection by the government.
	\item \textbf{Expenditure Policy}: Policies that directly impact the government's expenditure.
	\item \textbf{Non-Operating Income Policy}: Policies that control the government's non-operating income.
	\item \textbf{Irregular Items Policy}: Policies that control the government's irregular cash flow items.
	\item \textbf{Auxiliary Policy}: Policies introduced by the government with the purpose of stabilizing the economy, but that do not directly impact the government's financial position.
\end{itemize}
The notable exception that we make to the proposed structure above are the policies that primarily target the imports and exports, as these policies we cover in section \ref{sec:InternationalTradePolicies}.\par
\subsection{Revenue Policy}
\emph{Revenue Policy} is a set of atomic policies that target an increase or decrease in government revenue collection. We will preserve the structure that we have introduced in section \ref{subsec:GovernmentIncomeStatement} and will split the \emph{Revenue Policy} into \emph{Tax Revenue Policy} and \emph{Non-Tax Revenue Policy}, as a lot of tax-related policies have similar mechanisms of their implementation.
\subsubsection{Tax Revenue Policy}
In this section we look at different atomic policies that a government can introduce or remove from the economy, but first we would like to introduce a concept of a trait.\par
For different taxation channels defined in section \ref{subsec:GovernmentIncomeStatement}, there exist similarities in the way that they can be implemented. For instance, \emph{Personal Income Tax} may use a ladder system, where the tax rate is determined by the income band, but it may also have a flat rate that is constant for all income levels, or a fixed tax amount that has to be paid regardless of the base level of income. The same approach can be used for other taxes such as for \emph{Dividend Tax} or \emph{Capital Gains Tax}.\par
Hence, we introduce a mechanism that will simplify listing all the atomic policies with similar properties. We will be using traits throughout the paper, but the reason we define them as part of this section is because it is an abstraction that is best illustrated by an example.\par
\begin{definition}
	\textbf{Trait} is a property of a cash flow, economic participant, or financial instrument, implementation of which forms an atomic policy. A trait may have predefined parameters that are needed for its implementation, and it may itself be a categorical variable limited to a number of possible subtraits with their own parameters.
\end{definition}
In essence, a trait helps in defining atomic policies as the government will have to make a decision on what traits to use and what paremeters to set for each trait implementation (e.g., a government may choose to implement a flat rate tax as a trait for \emph{Personal Income Tax}, but it also needs to set the tax rate as a parameter).\par
The \emph{Personal Income Tax} example is interesting because there could be multiple traits that could be implemented for a tax category, but they might be mutually exclusive (i.e., the trait behaves as the categorical variable). As in the case above, \emph{Personal Income Tax} cannot have a progressive tax system and a flat rate at the same time. To deal with mutual exclusivity, we define an overarching trait and inside of it we introduce different mutually exclusive subtraits that could be implemented for a specific category. For the example above we introducee \emph{Tax Calculation Type}, which describes different ways for calculating the tax, and introduce the subtraits such as \emph{Proportional Tax} and \emph{Pregoressive Tax} that are different mutually exclusive subtraits that can be implemented for a given tax category.\par
For taxes, we introduce the following traits and their parameters (note that parameters are defined in the square brackets after the trait's name):
\begin{itemize}
	\item \textbf{Tax Calculation Type}: Type of calculation performed to compute the tax amount levied.
	\begin{itemize}
		\item \textbf{Proportional Tax} [tax rate]: Fixed tax rate applied regardless of other variables.
		\item \textbf{Progressive Tax} [ladder of tax rates]: Taxation regime under which a progressive series of bands are created with consecutive bands having a higher tax rate depending on a desired variable (e.g., income, wealth).
		\item \textbf{Regressive Tax} [ladder of tax rates]: Taxation regime under which a progressive series of bands are created with consecutive bands having a lower tax rate depending on a desired variable (e.g., income, wealth).
		\item \textbf{Lump-Sum Tax} [tax amount]: Fixed amount of tax that has to be paid by the entity to the government.
	\end{itemize}
	\item \textbf{Tax Base}: Base value to compute tax from.
	\begin{itemize}
		\item \textbf{Ad-Valorem Tax} [specified good or service]: Tax base is the value of a good or service.
		\item \textbf{Per-Unit Tax} [specified good or service]: Tax base is the quantity of the underlying being taxed.
	\end{itemize}
	\item \textbf{Tax Payment Type}: The moment at which the tax is paid.
	\begin{itemize}
		\item \textbf{Period Payment} [period]: Tax is aggregated over a certain period before being paid.
		\item \textbf{Automatic Payment}: Tax is paid at the time of the transaction.
	\end{itemize}
	\item \textbf{Tax Evasion Penalty}: Penalty charged in the case of tax not being paid within the dedicated window of time.
	\begin{itemize}
		\item \textbf{Tax Evasion Rate} [rate]: Fixed rate applied as a penalty for tax evasion.
		\item \textbf{Tax Evasion Ladder} [ladder of rates]: Ladder of rates applied as penalties for tax evasion.
		\item \textbf{Fixed Penalty} [fixed amount]: Fixed amount paid as a penalty for tax evasion.
	\end{itemize}
	\item \textbf{Allowance} [amount]: Fixed amount that is tax free, before further tax calculations are applied.
	\item \textbf{Abatement}: Temporary reduction or elimination of tax.
	\begin{itemize}
		\item \textbf{Abatement Rate} [period, rate]: Temporary reduction of tax by the abatement rate.
		\item \textbf{Abatement Ladder} [period, ladder of rates]: Temporary reduction of tax based on the ladder of rates.
		\item \textbf{Abatement Amount} [period, amount]: Temporary reduction of tax by the fixed amount.
	\end{itemize}
	\item \textbf{Exemption} [exemption condition, reduction amount]: Reduction or removal of a liability to make a compulsary payment, based on a certain condition, that would otherwise be imposed by a tax authority.
	\begin{itemize}
		\item \textbf{Tax Exemption Rate} [exemption condition, rate]: Reduction of tax liability by a certain rate based on a certain condition.
		\item \textbf{Tax Exemption Ladder} [exemption condition, ladder of rates]: Reduction of tax liability by the appropriate rate from the ladder based on a certain condition.
		\item \textbf{Tax Exemption Amount} [exemption condition, amount]: Reduction of tax liability by a fixed amount based on a certain condition.
	\end{itemize}
	\item \textbf{Negative Tax} [condition, amount]: Payment made by the government to taxpayer if the taxpayer satisfies certain requirement.
	\item \textbf{Tax Credit} [condition, amount]: Tax incentive which allows certain taxpayers to subtract the amount of credit they have accrued from the total tax.
\end{itemize}
Note, that the implementation of traits defined above is an atomic policy, as the government decides on exactly what trait will be applied with the specified parameter and on what tax category. So we now we can simplify the definitions of the atomic policies under the \emph{Tax Revenue Policy} by creating a table of tax categories and the traits that they can implement. However, before doing so we would also like to mention the other advantage of using traits, which is that they are extendable. If we were to add a new trait to the list above, all we need to do to update the list of policies is to define what tax categories can implement thast trait. this is a convenient tool for extending the existing taxonomy in the future and it ensures that we cover all possible combinations of tax categories and their traits.\par
Now we construct tables of trait implementations, which are the table that describe what traits can be implemented for what tax categories. We could use one big table with all tax categories and their traits, but for the purpose of the paper we decided to split it into multiple tables, each realting to their categorization in section \ref{subsec:GovernmentIncomeStatement}(e.g., \emph{Income Tax}, \emph{Property Tax}, etc.).\par
Note that some tax categories may have their own parameters, aside from the parameters required by the traits. We list the tax category parameters in square brackets under the tax category name. Appendix \ref{appendix:FiscalPolicy} contains tables \ref{table:FiscalPolicy:RevenuePolicy:TaxRevenuePolicy:IncomeTax}, \ref{table:FiscalPolicy:RevenuePolicy:TaxRevenuePolicy:PropertyTax}, \ref{table:FiscalPolicy:RevenuePolicy:TaxRevenuePolicy:SalesTax} and \ref{table:FiscalPolicy:RevenuePolicy:TaxRevenuePolicy:OtherTaxCategories}, that we have constructed (note that the rows in yellow are relevant to section \ref{sec:InternationalTradePolicies}). The checkmark points out that a certain tax category can implement a specific trait. Therefore, every checkmark in these tables corresponds to an atomic policy that a government can implement or remove.\par
There are also some tax-related policies that are unique in their definition, these are:
\begin{itemize}
	\item \textbf{Tax-Free Investment Accounts}: Tax incentive which allows taxpayers to deposit money in investment or saving accounts where all gains are tax-free.
	\item \textbf{Tax Amnesty}: Volunarily Disclose of tax liability and the payment of outstanding tax that does not invoke tax evasion penalties.
	\item \textbf{Tax-Free Shopping}: Visitors to a country can get sales tax refund.
\end{itemize}
This sums up the tax-related atomic policies that we initially introduce in our framework, but to conclude we would like to comment on the methodology that went into assembling \emph{Tax Revenue Policies}. We started by categorizing the tax-related policies based on their thematic link with each other. For instance, \emph{Personal Income Tax} and \emph{Capital Gain Tax} are both taxes that are charged on some form of income, and therefore, we placed them underneath the \emph{Income Tax} category. However, there are many specialized taxes, like alcohol tax, fat tax or window tax. To deal with these, we either tried to create a generalized category that would cover these (e.g., we introduced \emph{Pigouvian Good Tax} category to account for alcohol tax, tobacco tax, fat tax and etc.), or we added them to the taxonomy because of their unique effects (e.g., \emph{Carucage} is unique in its effect), or we ignored the tax policy if it was too specific and outdated (e.g., we excluded window tax, which is a tax on ther number of windows in the house, as it is very outdated and is unlikely to be of importance to current economic theory). We leave the cases of very ``specialized'' policies for the future classification.\par
\subsubsection{Non-Tax Revenue Policy}
This section is concerned with policies regaring non-tax revenue collection and corresponds to the policy decisions regarding the \emph{Non-Tax Revenue} in section \ref{subsec:GovernmentIncomeStatement}. The \emph{Non-Tax Revenue} items are more disjointed than the ones in the \emph{Tax Revenue} section, so we can adopt the ``traits''-based approach only for some of them.\par
We can create subcategorize \emph{Non-Tax Revenue Policy} into policies impacting the \emph{Government's Goods and Services}, and into other atomic policies. To cover the atomic policies for \emph{Government's Goods and Services} we introduce the following traits:
\begin{itemize}
	\item \textbf{Price Level} [good or service, amount]: Price level that is set for a specific good or service provided by the state-owned enterprise.
\end{itemize}
This allows us to create a table of different \emph{Non-Tax Revenue} items that can adopt these traits, and this table is presented in appendix \ref{appendix:FiscalPolicy} as table \ref{table:FiscalPolicy:RevenuePolicy:NonTaxRevenuePolicy:GovernmentGoodsAndServices}.\par
Other atomic policies that the government can use to impact the \emph{Non-Tax Revenue} are:
\begin{itemize}
	\item \textbf{Intragovernmental Aid Provision} [condition, government's compartment, amount]: Amount of aid that the specific government compartment has to provide to the government given a specific condition.
	\item \textbf{Equalization Payment Amount} [subnational government, amount] Amount that will paid by the federal government to the subnational government as the equalization payment.
	\item \textbf{Seigniorage Incentive}: Issuance of collectable cash and coins to incentivize the general public into exchanging physical money at the mint.
	\item \textbf{International Seigniorage} [international goods and services, amount]: Amount of physical money to be issued that will be spent on purchasing foreign goods and services.
	\item \textbf{International Debt Issuance} [conditions, amount]: Amount of sovereign debt to issue to foreign creditors based on the specified conditions.
	\item \textbf{Domestic Debt Issuance} [conditions, amount]: Amount of soveriegn debt to issue to domestic creditors based on the specified conditions.
	\item \textbf{Equity Issuance} [amount]: Amount of sovereign equity to be issued and sold.
\end{itemize}
With this list we conclude the atomic policies that we introduce for the \emph{Non-Tax Revenue} items of the income statement in section \ref{subsec:GovernmentIncomeStatement}.\par
\subsection{Expenditure Policy}
\emph{Expenditure Policy} covers the government spending policies, and corresponds to the \emph{Expenses} items in section \ref{subsec:GovernmentIncomeStatement}. For \emph{Other Expenses} we can use define traits as these expenses bahave in a similar way to each other. The traits we introduce for the items in \emph{Other Expenses} are defined as:
\begin{itemize}
	\item \textbf{Payment Size} [agent, amount]: Amount that will be issued to the specified economic agent.
\end{itemize}
The table with \emph{Other Expense} categories is presented in appendix \ref{appendix:FiscalPolicy} as table \ref{table:FiscalPolicy:ExpenditurePolicy:OtherExpenses}. Notice that in this table we have ommited \emph{Negative Tax Payments} as it is determined by the implementation of \emph{Negative Tax} trait to for a specific tax category, so it not an \emph{Expenditure Policy}, but the impact of the \emph{Tax Revenue Policy}. Also, the \emph{Import Subsidization} and \emph{Export Subsidization} are coloured in yellow and we will cover them as part of \emph{International Trade Policy} in section \ref{sec:InternationalTradePolicies}.\par
All other atomic policies that a government may apply to impact its expenditure are listed below:
\begin{itemize}
	\item \textbf{Wage Level} [job, wage]: Wage set for a specific job provided by the state-owned organizations.
	\item \textbf{Fixed Asset Replacement} [fixed asset, cost]: Cost of replacing a specified fixed asset owned by the government.
\end{itemize}
Notice that we omitted the purchase of fixed assets from the \emph{Expenditure Policy} as we will cover this in the section on \emph{Irregular Items Policy}, since is is an irregular commitment on the government's income statement.\par
\subsection{Non-Operating Income Policy}
In this section we cover the policies that relate to the \emph{Non-Operating Income} items of the government's income statement from section \ref{subsec:GovernmentIncomeStatement}. The list of atomic policies for this subcategory is:
\begin{itemize}
	\item \textbf{Sovereign Wealth Fund Investment} [amount]: Amount that the government is placing into a sovereign wealth fund.
	\item \textbf{Sovereign Wealth Fund Withdrawal} [amount]: Amount that the government withdraws from a sovereign wealth fund.
\end{itemize}
\subsection{Irregular Items Policy}
\emph{Irregular Items Policy} covers a government's policies regarding the items in the \emph{Irregular Items} in section \ref{subsec:GovernmentIncomeStatement}, with the m,olecular policies listed below:
\begin{itemize}
	\item \textbf{Reparation Payment} [amount]: Amount that the government decides to pay in reparations.
	\item \textbf{Sale of Fixed Asset} [fixed asset, price]: Sale of the specified fixed asset at the specified price.
	\item \textbf{Purchase of Fixed Asset} [fixed asset, price]: Purchase of the specified fixed asset at the specified price.
\end{itemize}
\subsection{Auxiliary Policy}
Lastly, we have \emph{Auxuliary Policy}, which is a collection of policies that do not fit under other \emph{Fiscal Policy} subcategories. The atomic policies inside the \emph{Auxiliary Policy} are:
\begin{itemize}
	\item \textbf{Budget Balance Requirement} [upper bound, lower bound]: Upper or lower bound that the government's budget balance should satisfy.
	\item \textbf{Interest Rate Ceiling} [ceiling]: Ceiling on the interest rate charged by loan institutions.
	\item \textbf{Minimum Wage} [minimum wage]: Minimum wage requirement for the employers to pay to employees.
	\item \textbf{Maximum Hours} [maximum hours]: Maximum working hours that an employee can work for for an employer.
	\item \textbf{Regulation of Working Conditions} [industry, conditions]: Sanitation, inspections and other conditions of working facilities where employees work.
	\item \textbf{Labour Unionization}: Decision on whether to allow labour force to create entities that act as intermediarieas in negotiations with businesses to improve the working conditions for the labour force.
	\item \textbf{Retirement Age} [age]: Age at which a person is expected or required to cease work.
	\item \textbf{Defined Benefit Plan} [period, amount]: Defined periodic payments made in retirement by the sponsor of the scheme (e.g., employer) for each period.
	\item \textbf{Defined Contribution Plan} [period, amount]: Defined amounts are paid in during working life by the sponsor of the scheme (e.g., employer) for each period.
\end{itemize}

\section{Monetary Policy}
Monetary auhthority adopts \emph{Monetary Policy} to stabilize an economy, but conversely to the \emph{Fiscal Policy} adopted by the sovereign, monetary authority may adopt any benchmarks for the performance of their policy set. The most common metrics that money authorities use to measure the success of their policies are interest rates on different types of debt, spot price of a currency pair, growth of money supply, spot price of an  asset or commodity (e.g., gold), or GDP.\par
In practice, monetary authority is often independent from the government and has an entirely different set of tools to impact the economy and markets. Because of this we can often treat monetary authority as having its own income statement, and therefore, different financial operations that it can perform to achieve its set goals. We have presented the income statement for monetary authority in section \ref{subsec:MonetaryAuthorityIncomeStatement}, and now we will categorize the atomic policies that the monetary authority commonly uses.\par
However, from the monetary authority's income statement we see that most items on itconstitute what is known as the \emph{Open Market Operations}, and it is a commonly used term in economics. To better integrate the framework we propose with the current econoics theory, we will not categorize \emph{Monetary Policies} based of the monetary authority's income statement as we have done with the \emph{Fiscal Policy}, but rather propose a different structure that aim to preserve the existing subcategorization of \emph{Monetary Policy}:
\begin{itemize}
	\item \textbf{Open Market Operations Policy}: Policies regarding the purchase and sale of securities in the open market by the monetary authority.
	\item \textbf{Debt and Credit Policy}: Policies regarding creation of debt and credit by the monetary authority.
	\item \textbf{Financial Markets Policy}: Requirements that a monetary authority imposes on the financial markets.
	\item \textbf{Operation Policy}: Policies regarding the way the monetary authority is set up internally.
	\item \textbf{Communications Policy}: Policies that monetary authority uses to communicate its policy-making decisions.
\end{itemize}
\subsection{Open Market Operations Policy}
\emph{Open Market Operations Policy} is a subcategory of \emph{Monetary Policy} that includes all atomic policies regarding the purchase and sale of securities in the open market by the monetary authority. The reasons for a monetary authority to engage in these operations depends on the strategy and the targets outlined by the central bank, but any decision to purchase or sell a security we consider as an atomic policy.\par
Since all \emph{Open Market Operations} are similar to one another from perspective of purchasing and selling securities, we again can create traits that will help us define atomic policies for each component of \emph{Open Market Operations}.
\begin{itemize}
	\item \textbf{Security Purchase} [security, amount]: Purchase of the given amount of a certain security in the open market.
	\item \textbf{Security Sale} [security, amount]: Sale of the given amount of a certain security in the open market.
\end{itemize}
With these trait definitions we compose the table that reflects what traits each component of \emph{Open Market Operations} can implement, and therefore, facilitate an atomic policy. The table is presented in appendix \ref{appendix:MonetaryPolicy} as table \ref{table:MonetaryPolicy:OpenMarketOperationsPolicy}.\par
Notice that we did not make a destinction between central bank purchasing different types of contracts, for example to purchase foreign currency the central bank can use swaps, forward agreements and etc. We assume that this information is supplied into the ``security'' variable of the traits defined above.
\subsection{Debt and Credit Policy}
The \emph{Debt and Credit Policy} covers the polcies regarding loans that account holders have with the central bank, as well as the loans taken by the central bank itself. We will not cover the loan requirements that a central bank may introduce as we cover them as part of the \emph{Financial Markets Policy}, and we will also not cover the \emph{Repurchase Agreements} as we consider the trading of them to be part of the \emph{Open Market Operations Policy}.\par
For the loans with the monetary authority we define the following traits:
\begin{itemize}
	\item \textbf{Credit Issuance} [maturity, principal, deposit rate]: Issuance of a loan by the monetary authority.
	\item \textbf{Credit Quota} [condition, limit]: Quota on the amount of money that the monetary authority issues to an entity based on a specified condition.
	\item \textbf{Credit Embargo} [condition]: Ban on credit issuance by the monetary authority to an entity based on a specific condition.
	\item \textbf{Differential Interest Rate} [rate ladder with conditions]: Different deposit rates on loans with the monetary authority to different entities based on the conditions specified for each deposit rate.
	\item \textbf{Debt Issuance} [maturity, principal, deposit rate]: Issuance of debt by the monetary authority.
	\item \textbf{Debt Default} [debt]: Decision of the monetary authority to default on a specified debt.
\end{itemize}
The table \ref{table:MonetaryPolicy:DebtAndCreditPolicy} in appendix \ref{appendix:MonetaryPolicy} presents the traits implementable for the loans. The atomic policies where the traits do not apply are listed below:
\begin{itemize}
	\item \textbf{Helicopter Money} [amount]: Process of central bank creating money (without assets as a counterpart) in its balance sheet and issuing it directly to households.
\end{itemize}
Notice that \emph{Helicopter Money} is a hypothetical policy that has never been implemented in practice, but we still proceed with adding it to the taxonomy tree for future reference.
\subsection{Financial Markets Policy}
In this section we describe the limitations and requirements that a monetary authority can impose on the financial markets and conduct within them. The traits of the \emph{Financial Markets Policy} are:
\begin{itemize}
	\item \textbf{Administered Rate} [rate]: Rate that is manually set by the monetary authority for a specific financing market.
	\item \textbf{Rate Corridor} [lower bound, upper bound]: Lower and upper bounds that are set by the monetary authority to restrict a specific financing market.
	\item \textbf{Collateral Assets} [assets]: Assets that the monetary authority allows to be placed as a collateral.
	\item \textbf{Margin Requirement}: Minimum amount that the monetary authority requests to be placed as a collateral.
	\begin{itemize}
		\item \textbf{Fixed Margin} [amount]: Minimum fixed amount to be placed as a collateral.
		\item \textbf{Margin Ratio} [ratio]: Minimum ratio of collateral to the total value of the deal.
	\end{itemize}
\end{itemize}
With these traits, we have defined table \ref{table:MonetaryPolicy:FinancialMarketsPolicy} in appendix \ref{appendix:MonetaryPolicy} for different financial markets that implement these traits.
\subsection{Operation Policy}
The \emph{Operation Policy} encompases the requirements that the central bank places on itself in order to operate at ``higher'' efficiency or to satisfy legal requirements. The atomic policies in this subcategory differ in their properties so we cannot use ``traits'' to list all the atomic policies inside a table as we have done in previos sections. Instead we opt to list them directly:
\begin{itemize}
	\item \textbf{Reserve Requirements}: Regulation that sets the minimum amount that a commercial bank must hold in liquid assets.
	\begin{itemize}
		\item \textbf{Full-Reserve Banking}: System of banking where banks do not lend demand deposits and instead, only lend from time deposits.
		\item \textbf{Fractional Reserve Banking}: System of banking under which banks that take deposits from the public must keep only a part of their deposit liabilities in liquid assets as a reserve.
		\begin{itemize}
			\item \textbf{Fractional Reserve Requirement} [amount]: Minimum amount that a commercial bank must hold in liquid assets with the central bank.
		\end{itemize}
	\end{itemize}
	\item \textbf{Fixed Exchange Rate} [pivot asset]: System in which a currency's value is fixed or pegged by a monetary authority against the value of another currency, a basket of other currencies, or another measure of value (i.e., pivot asset), such as gold.
	\item \textbf{Exchange Requirements} [ratio]: Requirement for the certain ratio of the foreign exchange receipts (generally from exports) to be exchnaged for the local currency in order to influence money supply in the economy.
	\item \textbf{Discount Window Brokerage}: Monetary authority sets up a discount window to broker interbank repurchase agreements.
	\item \textbf{Private Debt Limit} [amount]: Limit on the amount of corporate debt that can issued in the economy.
\end{itemize}
\subsection{Communications Policy}
The \emph{Communications Policy} subcategory covers the channels through which the central bank communicates its policy-making decisions to the markets. We include \emph{Communications Policy} as a subcategory of \emph{Monetary Policy} because the way the \emph{Monetary Policy} (i.e., in this case a set of atomic policies) is communicated can have an impact on the market, and is often used by central banks to do just that. Therefore, we see all the communications made by the central bank as a policy and categorize the communication channels in the following way:
\begin{itemize}
	\item \textbf{Forward Guidance}: Communication about future course of monetary policy.
	\begin{itemize}
		\item \textbf{Odyssean Forward Guidance}: Explicit forward guidance expressed through communication of forecasts and future intentions.
		\item \textbf{Delphic Forward Guidance}: Implied forward guidance expressed through ``less-binding'' version of forward guidance, which may not provide specific details for future policy decisions.
	\end{itemize}
	\item \textbf{Non-Official Communication}: Communications and ``hints'' from the members of the central bank that are not made during an official press release of the central bank.
\end{itemize}

\section{International Trade Policies}
\label{sec:InternationalTradePolicies}
The \emph{International Trade Policy} is designed to regulate international trade that the country engages in, but the same policy can also be used to stabilize or boost the economy. It is important to note that the atomic policies listed in this section are the policies that primarily target the internation trade regulation, since there are policies that we have mentioned in the previos sections that can heavily impact exports and import, but their main target is not the regulation of trade, so we do not include them here.\par
For example, a monetary authority may significantly lower the marginal lending rate on its repurchase agreements making it more attractive to foreign banks, which in turn results in higher demand for domestic currency. This can lead to improved financial conditions for the production sectors that sell expensive goods and services (e.g., automotive, aerospace and defence, luxury), but at the same time worsens the conditions for sectors that benefit from lower exchange rate (e.g., tourism). Clearly, that can offset the balance of exports and imports, yet the primary goal of this policy may have been to support domestic banks with the cheaper cost of liquidity.\par
We break down \emph{International Trade Policies} into \emph{Financial Trade Policy} and \emph{Legal Trade Policy}.\par
\subsection{Financial Trade Policy}
The \emph{Financial Trade Policy} encompases the tools at government's disposal where the deterrent or incentive is the  monetary ``penalty'' or ``reward''. The objective of these policies is usually to introduce a soft barrier (i.e., motivated parties can overcome these requirements at a cost) to entry for foreign businesses as well as the support of the domestic production.\par
Despite the fact that the atomic policies in this subcategory target the import and export, the mechanisms of their implementation are similar to some other policies that we have covered. In this sense, \emph{Tariff} and its traits are similar to the rest of the tax-related categories, and its trait implementations are defined in table \ref{table:FiscalPolicy:RevenuePolicy:TaxRevenuePolicy:OtherTaxCategories} (i.e., highlighted in yellow). Similarly, \emph{Import Subsidization} and \emph{Export Subsidization} are similar to other subsidization categories, and their trait implemantations are listed in table \ref{table:FiscalPolicy:ExpenditurePolicy:OtherExpenses} (i.e., highlighted in yellow).
\subsection{Legal Trade Policy}
The \emph{Legal Trade Policy} describes the limits and requirements that a government may implement to restrict the access of the country to foreign products and of foreign consumers to domestic products by introducing a hard barrier (i.e., motivated party will be persecuted for an attempt to bypass the requirements) to entry.
\begin{itemize}
	\item \textbf{Quantity Control}: Requirements imposed on the quantity of goods and services.
	\begin{itemize}
		\item \textbf{Import Quota} [good, total amount]: Allowance for the total number or amount of a specific imported good.
		\item \textbf{Export Quota} [good, total amount]: Allowance for the total number or amount of a specific exported good.
		\item \textbf{Import Embargo} [good or service]: Denial of import of a specific good or service.
		\item \textbf{Export Embargo} [good or service]: Denial of export of a specific good or service.
	\end{itemize}
	\item \textbf{Quality Control}: Requirements imposed on the quality of goods and services.
	\begin{itemize}
		\item \textbf{Domestic Standards} [good or service, conditions]: Requirements  and specifications that specific goods have to satisfy in order to be imported into the country (e.g., sanitary requirements).
	\end{itemize}
	\item \textbf{Anti-Dumping}: Legal actions imposed against parties that perform dumping in the domestic economy.
\end{itemize}

\section{Additional Features}
Having introduced the concept of a trait, we were able to define molecular policies and group them under in a policy taxonomy tree as demonstrated in Appendix \ref{appendix:TaxonomyRoot}. We mentioned the other benefit of traits in section \ref{sec:FiscalPolicy}, which is the extendability of this approach. If a new trait (or policy category) is introduced, we can add a column (row) to the tables in the appendicies to immediately gauge what other policy categories could implement this trait (what other traits could be implemented for this policy category). This allows for an exhaustive mapping between traits and policy categories.\par
The other advantage of using traits to define atomic policies is that the implementable traits can tell us how similar or different certain policy categories are. In this section we woul like to briefly demonstrate some of the methods that we have used to create visualizations of our policy taxonomy and the relationship between all policy groups that we have defined in this paper (all figures are provided in Appendix \ref{appendix:AdditionalFeatures}).\par
We start by creating a table where the columns are all the traits that we have defined across the paper, while the rows represent different policy categories (e.g., cash flows, economic participants, financial instruments and etc. that implement the traits that we have defined.). Each cell in this table is a boolean, and contains either ``True'' or ``False'', depending on whether a given policy category implements the trait from the column the cell is in. The resulting table will resemble the tables we have presented in appendices, but it includes all traits and policy categories, and replaces tickmarks with ``True'', while filling the blanks with ``False''. We also add policy categories that do not implement any traits, such as \emph{Maximum Wage} or \emph{Helicopter Money}, and set all of them not to implement any traits (i.e., all cells are set to ``False'').\par
\begin{figure}[htpb]
\begin{center}
\rotatebox{90}{
\includegraphics[scale=0.25]{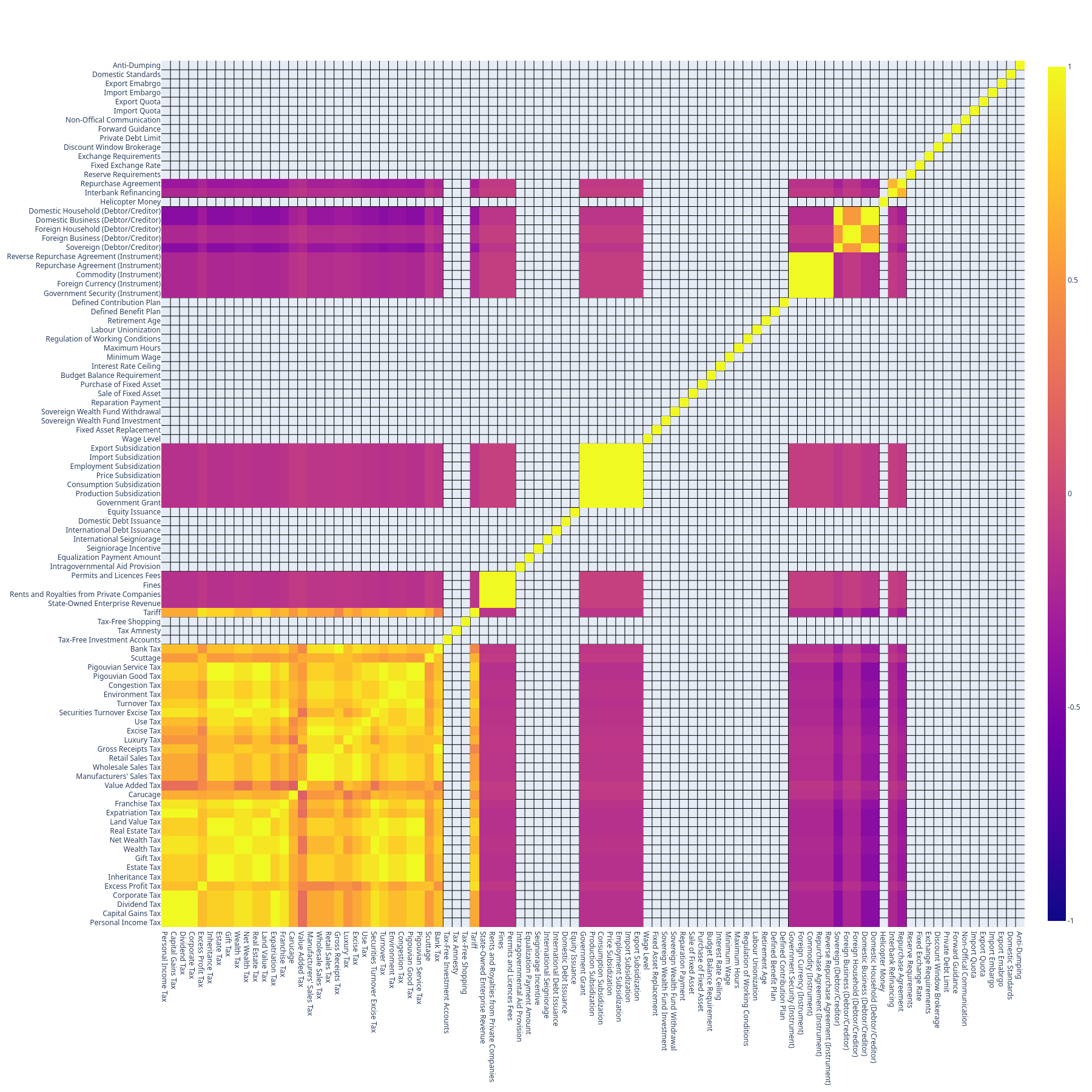}
}
\end{center}
\caption{Pearson correlation between policy categories based on what traits they implement}
\label{figure:Corrall}
\end{figure}

The reason for creating such a table is that now we can compare the how similar the policy categories are based on their trait implementations, as for each policy category we now have a signal series (i.e., a vector $S\in\mathbb{R}^{23}$, as we have 23 traits in total so far) that we can use to produce correlations between policy categories. For instance, we can compute Pearson correlation coefficients between policy categories to obtain the correlation matrix on Figure \ref{figure:Corrall}. Note that there are a lot of correlations that have not been computed (blue-grey coloured cells) since some signal series are constant (i.e., they have 23 repetitive values), which means their standard deviation is 0, and makes it impossible to compute their correlation with signal series of other policy categories.\par
Aside from finding correlations between the policy categories, we can employ these signal series to construct a minimum-spanning tree to visualize the relationship between the policy categories. For the minimum-spanning tree we define policy categories as the nodes of the graph, and we use euclidean distance to measure the distance between each policy category. These distances are the weights of the edges between the policy category nodess, and together they form a distance matrix which tells us how far apart a policy category is from other policy categories. Next, we employ Kruskal's algorithm to prune the distance matrix, so that the resulting distance matrix will only have the distances between the nodes (i.e., policy categories) that are adjacent in the minum-spanning tree, while all other distances in the distance matrix are set to ``None''. Finally we convert the resulting distance matrix to the adjacency matrix where the two nodes, $i$ and $j$, that are connected in the minimum-spanning tree have the coefficient of $a_{ij}=1$, while the nodes that are not connected have the coefficient $a_{ij}=0$.\par
\begin{figure}[htpb]
\begin{center}
\rotatebox{90}{
\includegraphics[scale=0.3]{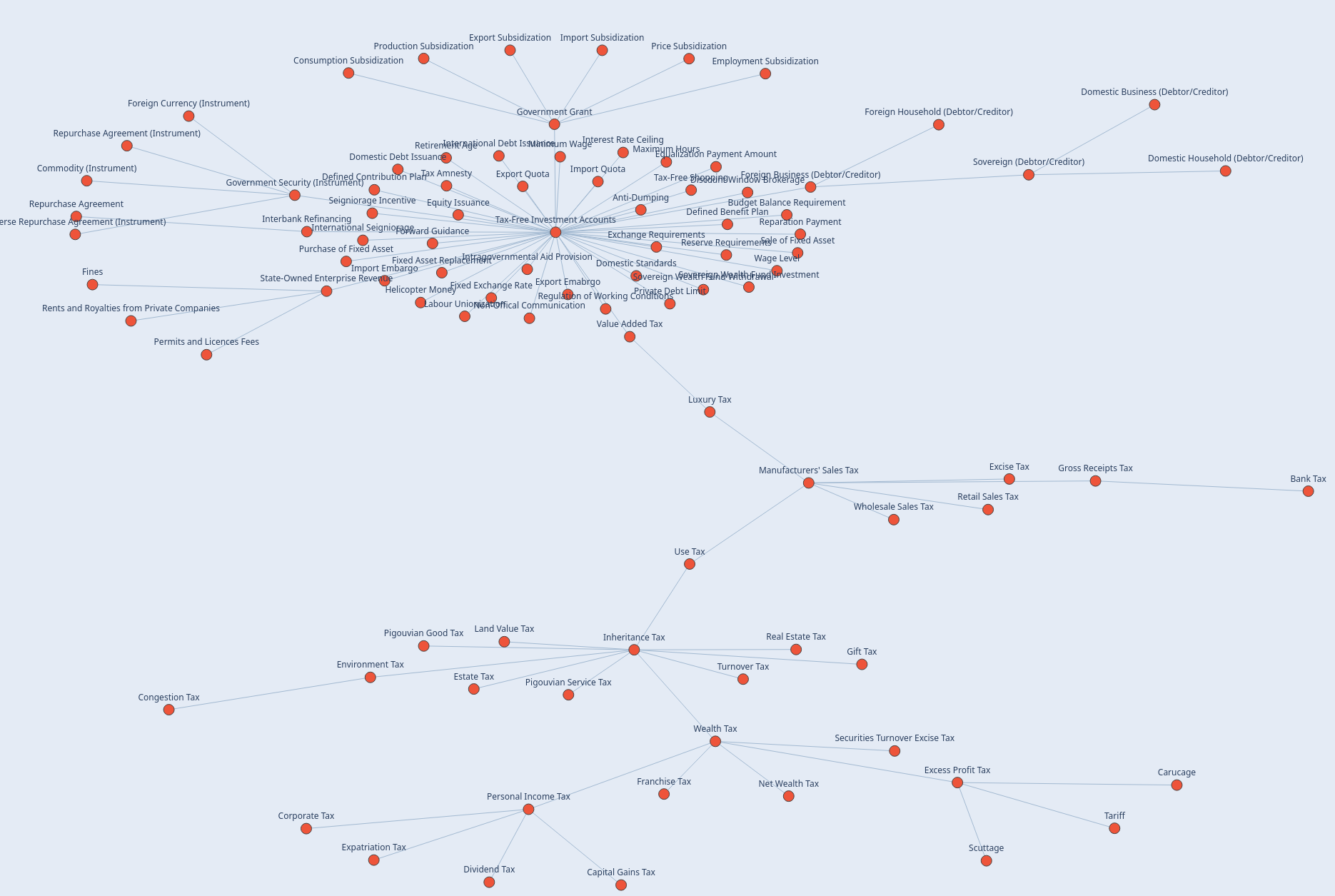}
}
\end{center}
\caption{Minimum-Spanning Tree of atomic policies (Includes policy categories that do not implement any traits)}
\label{figure:MSTAll}
\end{figure}
\begin{figure}[htpb]
\begin{center}
\rotatebox{90}{
\includegraphics[scale=0.3]{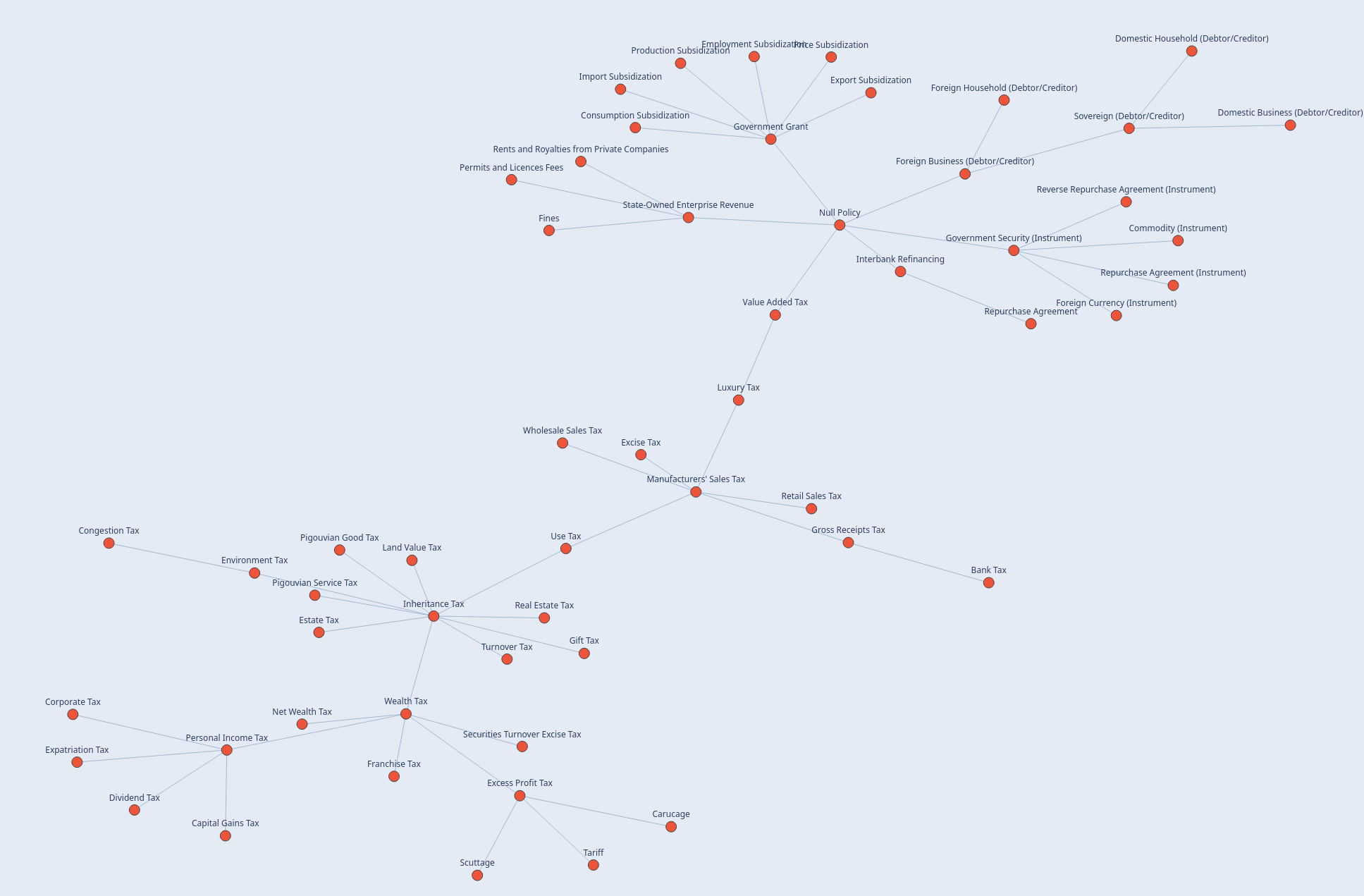}
}
\end{center}
\caption{Minimum-Spanning Tree of atomic policies (Replacing policy categories that do not implement any traits with ``Null Policy'')}
\label{figure:MSTOneNull}
\end{figure}
\begin{figure}[htpb]
\begin{center}
\rotatebox{90}{
\includegraphics[scale=0.3]{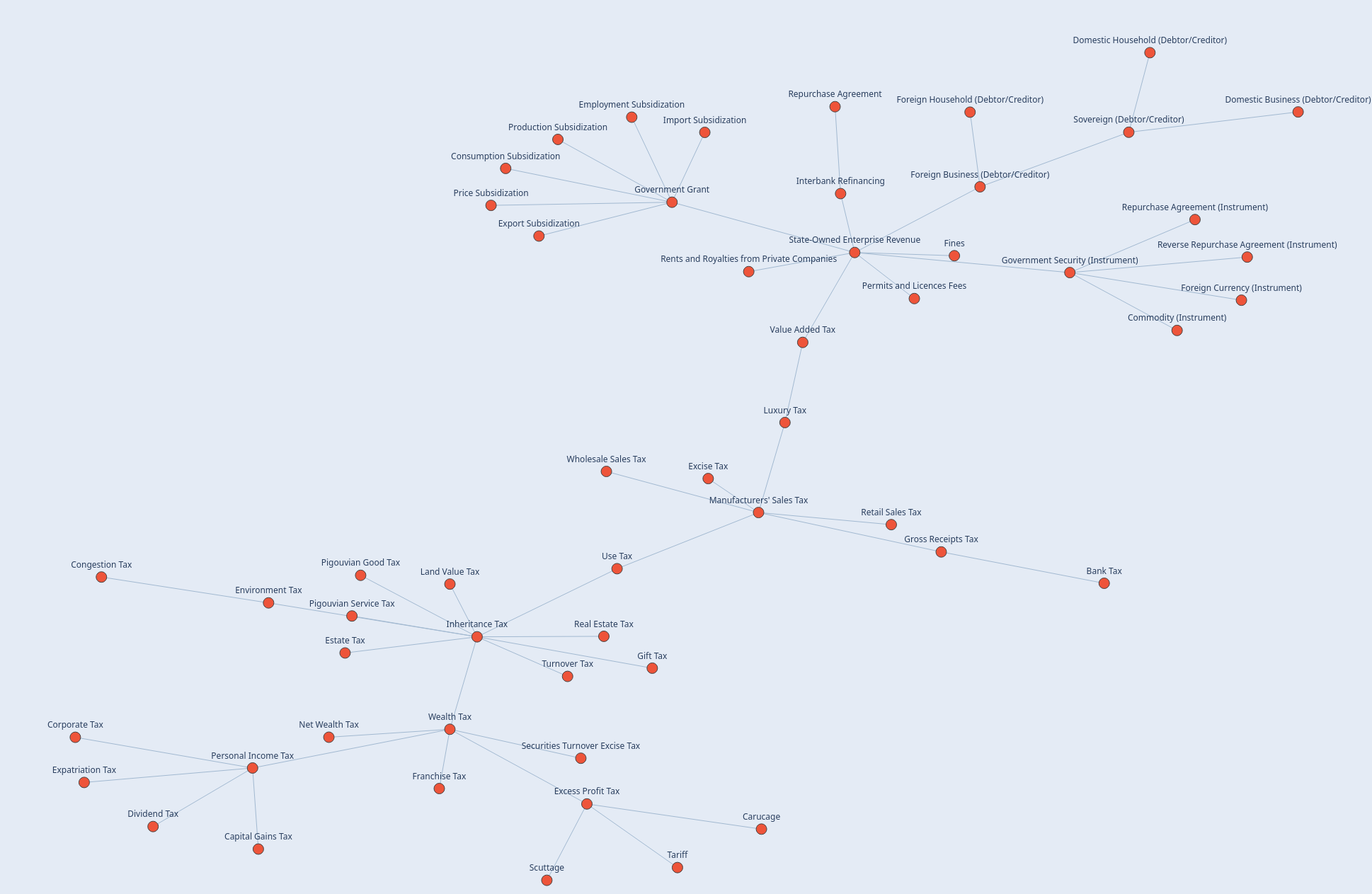}
}
\end{center}
\caption{Minimum-Spanning Tree of atomic policies (Excluding policy categories that do not implement any traits)}
\label{figure:MSTNotNull}
\end{figure}

If we perform the process described above, we obtain the the minimum-spanning tree presented on Figure \ref{figure:MSTAll}. Since we used the adjacency matrix to construct the graph, the distacne between the nodes is arbitrary and is set by the plotting software. Also, note that all the policy categories that do not implement any traits form a large cluster around the \emph{Tax-Free Investment Account} node since the distance between each of their signal series is 0. If we adjust our graph for that by replacing all policy categories that do not implement any traits with a single node called ``Null Policy'', we obtain the minimum-spanning tree on Figure \ref{figure:MSTOneNull}. If we were to filter out all policy categories that do not implement any traits, we obtain the graph in Figure \ref{figure:MSTNotNull}.\par
These analysis methods can be extended further: different distance metrics could be used; different algorithms can be used to find the minium-spanning tree; distance matrix can be used instead of adjacency matrix such that the distance between the nodes tells us how similar the policy categories are; or a closed graph can be defined where the distance matrix is filtered by some arbitrary condition. These analysis methods can be useful to visualise the relationship between the policy categories based on their implementation requirements, but they are solemnly based on the way we define the traits, so we do not provide the analysis in this section as the result of the paper, rather we demonstrate the convenient features of our framework for defining policy taxonomy using traits as the building blocks.\par

\section{Conclusion}
In summary, we presented commonly used economic policies and proposed an initial structure for their categorization, which can be expanded in the future. The objective of this paper is to present an initial taxonomy of economic policies in order to develop an exhaustive taxonomy in the future. The benefit of having a well-defined structure of potential policies is that it provides a list of actions that the regulators can use to construct economic strategies, and it presents a list of variables for economic models to use as an input or output. Having an exhaustive list of policies means that economic models that are developed using this framework can explicilty state what policies and variables have been omitted in a model.\par
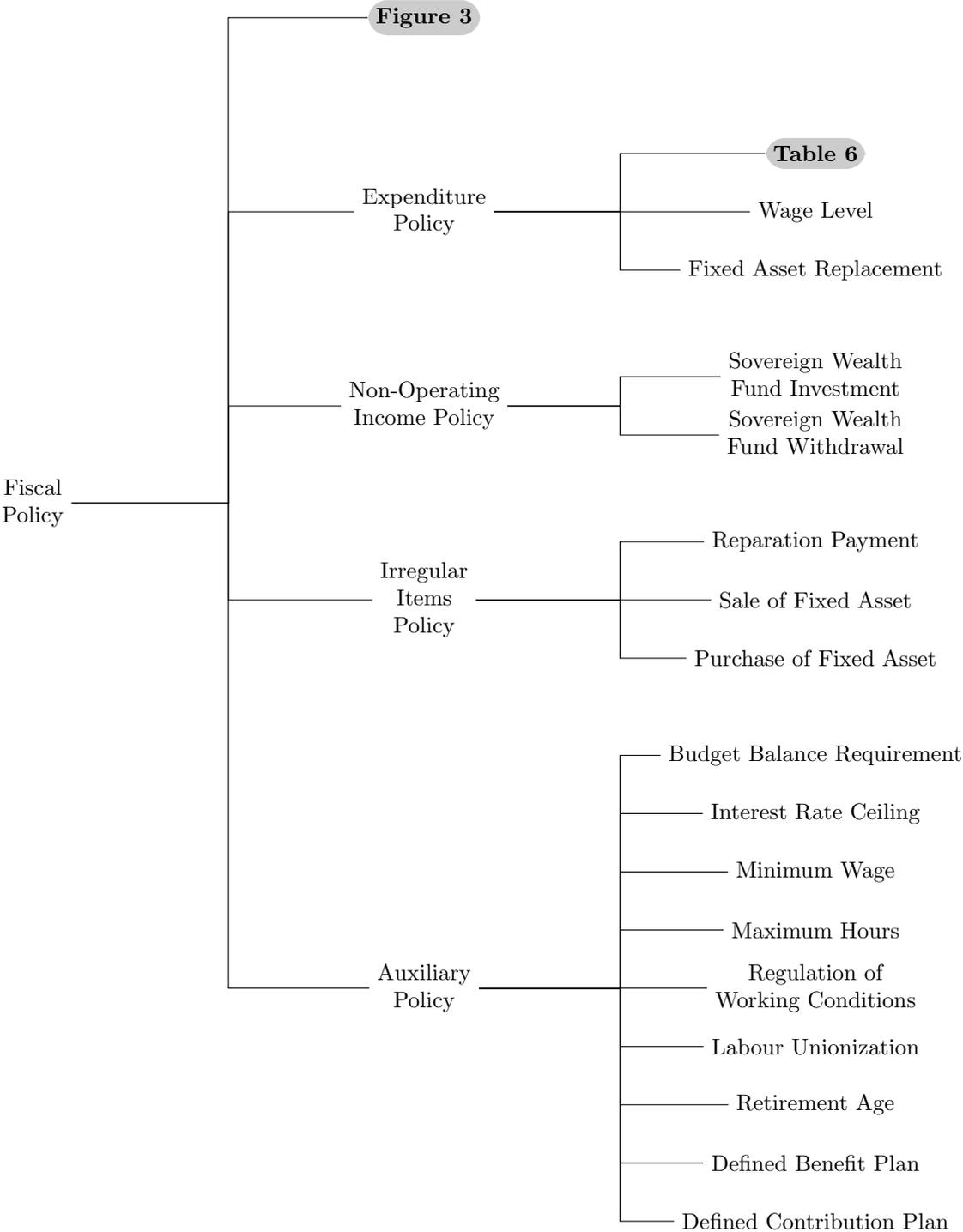
\begin{figure}[h!]
\begin{center}
\footnotesize
\begin{tikzpicture}[
	grow=right,
	level distance=60mm,
	level 1/.style={sibling distance=30mm},
	level 2/.style={sibling distance=9mm},
	edge from parent fork right,
]
	\node {\begin{tabular}{@{}c@{}}Fiscal \\ Policy\end{tabular}}
		child {
			node {\begin{tabular}{@{}c@{}}Auxiliary \\ Policy\end{tabular}}
				child {node {Defined Contribution Plan}}
				child {node {Defined Benefit Plan}}
				child {node {Retirement Age}}
				child {node {Labour Unionization}}
				child {node {\begin{tabular}{@{}c@{}}Regulation of \\ Working Conditions\end{tabular}}}
				child {node {Maximum Hours}}
				child {node {Minimum Wage}}
				child {node {Interest Rate Ceiling}}
				child {node {Budget Balance Requirement}}
		}
		child[missing] {node {}}
		child {
			node {\begin{tabular}{@{}c@{}c@{}}Irregular \\ Items \\ Policy\end{tabular}}
				child {node {Purchase of Fixed Asset}}
				child {node {Sale of Fixed Asset}}
				child {node {Reparation Payment}}
		}
		child {
			node {\begin{tabular}{@{}c@{}}Non-Operating \\ Income Policy\end{tabular}}
				child {node {\begin{tabular}{@{}c@{}}Sovereign Wealth \\ Fund Withdrawal\end{tabular}}}
				child {node {\begin{tabular}{@{}c@{}}Sovereign Wealth \\ Fund Investment\end{tabular}}}
		}
		child {
			node {\begin{tabular}{@{}c@{}}Expenditure \\ Policy\end{tabular}}
				child {node {Fixed Asset Replacement}}
				child {node {Wage Level}}
				child {node[rounded corners=0.3cm, fill=gray!40] {\textbf{Table \ref{table:FiscalPolicy:ExpenditurePolicy:OtherExpenses}}}}
		}
		child {
			node[rounded corners=0.3cm, fill=gray!40] {\textbf{Figure \ref{diagram:FiscalPolicy:RevenueOperationsPolicy}}}
		};
\end{tikzpicture}
\end{center}
\caption{Atomic policies in \emph{Fiscal Policy}}
\label{diagram:FiscalPolicy}
\end{figure}
\begin{figure}[h!]
\begin{center}
\footnotesize
\begin{tikzpicture}[
	grow=right,
	level distance=50mm,
	level 1/.style={sibling distance=100mm},
	level 2/.style={sibling distance=11mm},
	edge from parent fork right,
]
	\node {\begin{tabular}{@{}c@{}}Revenue \\ Operations Policy\end{tabular}}
		child {
			node {\begin{tabular}{@{}c@{}c@{}}Non-Tax \\ Revenue \\ Policy\end{tabular}}
				child {node {Equity Issuance}}
				child {node {Domestic Debt Issuance}}
				child {node {International Debt Issuance}}
				child {node {International Seigniorage}}
				child {node {Seigniorage Incentive}}
				child {node {\begin{tabular}{@{}c@{}}Equalization \\ Payment Amount\end{tabular}}}
				child {node {\begin{tabular}{@{}c@{}}Intragovernmental \\ Aid Provision\end{tabular}}}
				child {node[rounded corners=0.3cm, fill=gray!40] {\textbf{Table \ref{table:FiscalPolicy:RevenuePolicy:NonTaxRevenuePolicy:GovernmentGoodsAndServices}}}}
		}
		child {
			node {\begin{tabular}{@{}c@{}}Tax Revenue \\ Policy\end{tabular}}
				child {node {Tax-Free Shopping}}
				child {node {Tax Amnesty}}
				child {node {\begin{tabular}{@{}c@{}}Tax-Free \\ Investment \\ Accounts\end{tabular}}}
				child {node[rounded corners=0.3cm, fill=gray!40] {\textbf{Table \ref{table:FiscalPolicy:RevenuePolicy:TaxRevenuePolicy:OtherTaxCategories}}}}
				child {node[rounded corners=0.3cm, fill=gray!40] {\textbf{Table \ref{table:FiscalPolicy:RevenuePolicy:TaxRevenuePolicy:PropertyTax}}}}
				child {node[rounded corners=0.3cm, fill=gray!40] {\textbf{Table \ref{table:FiscalPolicy:RevenuePolicy:TaxRevenuePolicy:PropertyTax}}}}
				child {node[rounded corners=0.3cm, fill=gray!40] {\textbf{Table \ref{table:FiscalPolicy:RevenuePolicy:TaxRevenuePolicy:IncomeTax}}}}
		};
\end{tikzpicture}
\end{center}
\caption{Atomic policies in \emph{Revenue Operations Policy}}
\label{diagram:FiscalPolicy:RevenueOperationsPolicy}
\end{figure}
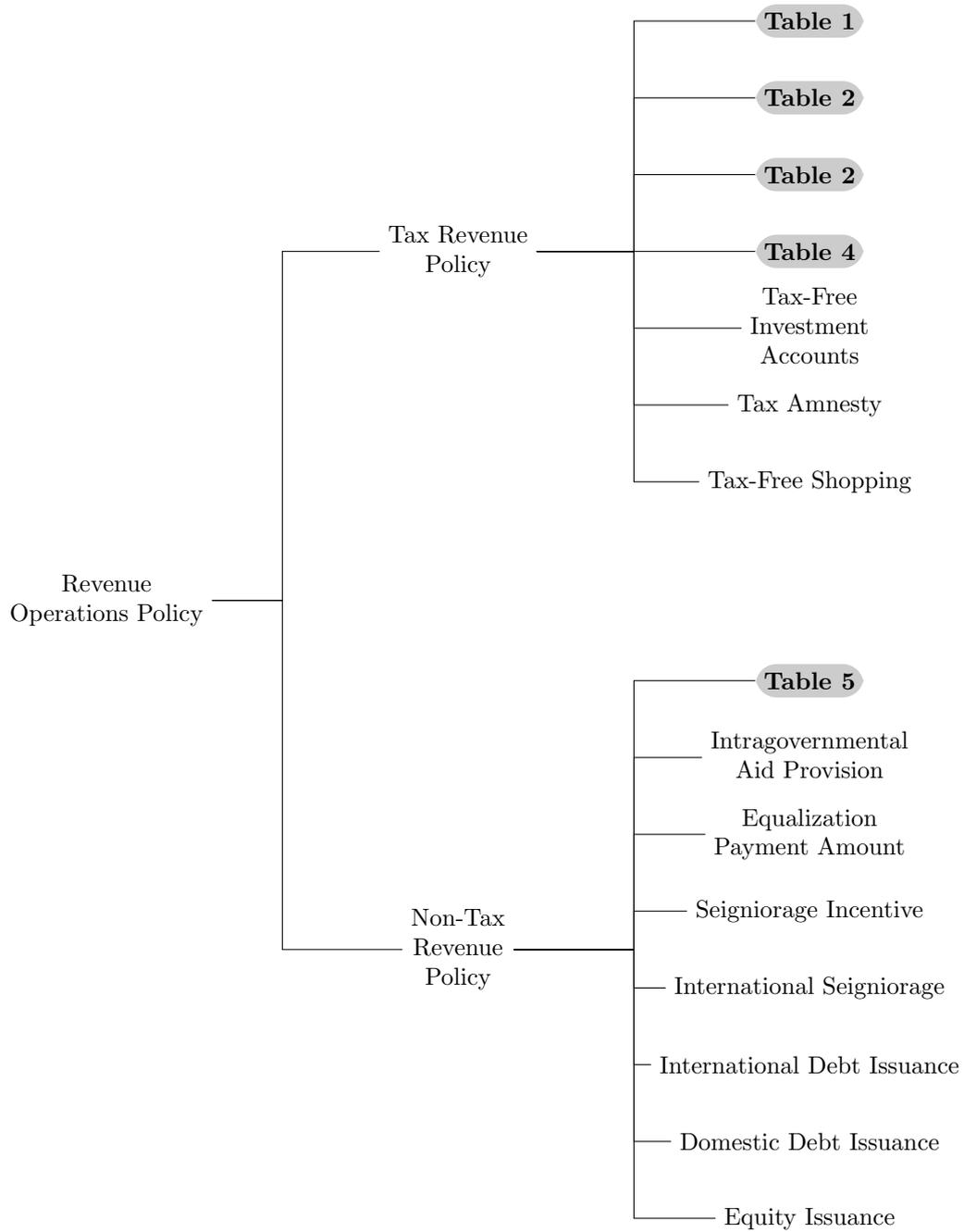
\begin{figure}[h!]
\begin{center}
\footnotesize
\begin{tikzpicture}[
	grow=right,
	level distance=35mm,
	level 1/.style={sibling distance=50mm},
	level 2/.style={sibling distance=15mm},
	level 3/.style={sibling distance=15mm},
	edge from parent fork right,
]
	\node {\begin{tabular}{@{}c@{}}Monetary \\ Policy\end{tabular}}
		child {
			node {\begin{tabular}{@{}c@{}}Communications \\ Policy\end{tabular}}
				child {
					node {\begin{tabular}{@{}c@{}}Non-Official\\ Communication\end{tabular}}
				}
				child {
					node {Forward Guidance}
					child {node {\begin{tabular}{@{}c@{}}Delphic Forward \\ Guidance\end{tabular}}}
					child {node {\begin{tabular}{@{}c@{}}Odyssean Forward \\ Guidance\end{tabular}}}
				}
		}
		child {
			node {\begin{tabular}{@{}c@{}}Operation \\ Policy\end{tabular}}
				child {node {\begin{tabular}{@{}c@{}}Private Debt \\ Limit\end{tabular}}}
				child {node {\begin{tabular}{@{}c@{}}Discount Window \\ Brokerage\end{tabular}}}
				child {node {\begin{tabular}{@{}c@{}}Exchange \\ Requirements\end{tabular}}}
				child {node {\begin{tabular}{@{}c@{}}Fixed \\ Exchanged Rate\end{tabular}}}
				child {
					node {\begin{tabular}{@{}c@{}}Reserve \\ Requirements\end{tabular}}
						child {
							node {\begin{tabular}{@{}c@{}c@{}}Fractional \\ Reserve \\ Banking\end{tabular}}
								child {node {\begin{tabular}{@{}c@{}c@{}}Fractional \\ Reserve \\ Requirement\end{tabular}}}
						}
						child {node {\begin{tabular}{@{}c@{}}Full-Reserve \\ Banking\end{tabular}}}
				}
		}
		child {
			node {\begin{tabular}{@{}c@{}}Financial Markets \\ Policy\end{tabular}}
				child {node[rounded corners=0.3cm, fill=gray!40] {\textbf{Table \ref{table:MonetaryPolicy:FinancialMarketsPolicy}}}}
		}		
		child {
			node {\begin{tabular}{@{}c@{}}Debt and Credit \\ Policy\end{tabular}}
				child {node[rounded corners=0.3cm, fill=gray!40] {\textbf{Table \ref{table:MonetaryPolicy:DebtAndCreditPolicy}}}}
				child {node {Helicopter Money}}
		}
		child {
			node {\begin{tabular}{@{}c@{}}Open Market \\ Operations Policy\end{tabular}}
				child {node[rounded corners=0.3cm, fill=gray!40] {\textbf{Table \ref{table:MonetaryPolicy:OpenMarketOperationsPolicy}}}}
		};
\end{tikzpicture}
\end{center}
\caption{Atomic policies in \emph{Monetary Policy}}
\label{diagram:MonetaryPolicy}
\end{figure}
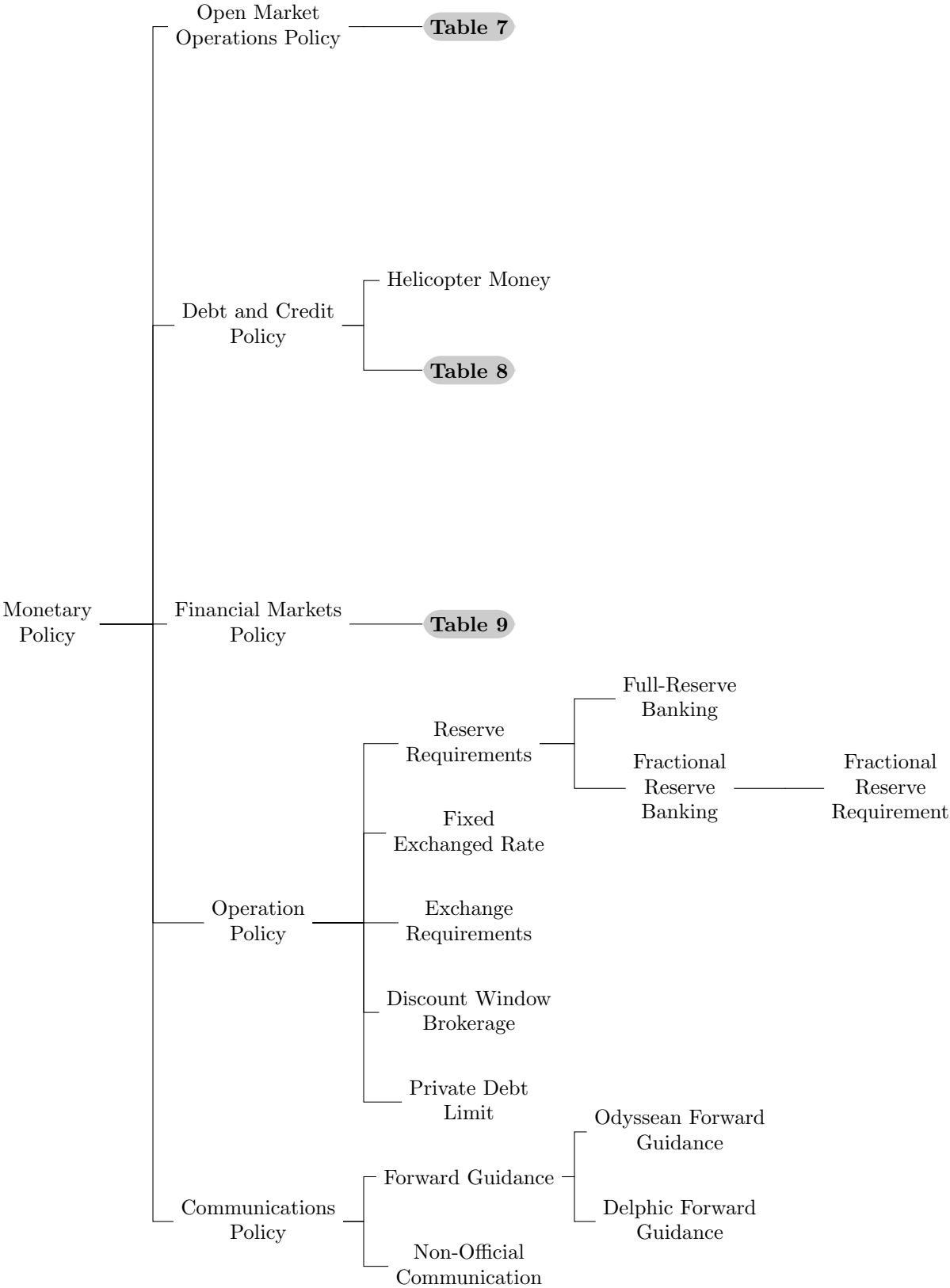
\begin{figure}[htpb]
\begin{center}
\footnotesize
\begin{tikzpicture}[
	grow=right,
	level distance=45mm,
	level 1/.style={sibling distance=70mm},
	level 2/.style={sibling distance=25mm},
	level 3/.style={sibling distance=9mm},
	edge from parent fork right,
]
	\node {\begin{tabular}{@{}c@{}}International \\ Trade Policy\end{tabular}}
		child {
			node {\begin{tabular}{@{}c@{}}Legal \\ Trade Policy\end{tabular}}
				child {node {Anti-Dumpting}}
				child {
					node {Quality Control}
						child {node {Domestic Standards}}						
				}
				child {
					node {Quantity Control}
						child {node {Export Embargo}}
						child {node {Import Embargo}}
						child {node {Export Quota}}
						child {node {Import Quota}}
				}
		}
		child {
			node {\begin{tabular}{@{}c@{}}Financial \\ Trade Policy\end{tabular}}
				child {
					node[rounded corners=0.3cm, fill=gray!40] {\textbf{Table \ref{table:FiscalPolicy:ExpenditurePolicy:OtherExpenses} (Yellow Rows)}}
				}
				child {
					node[rounded corners=0.3cm, fill=gray!40] {\textbf{Table \ref{table:FiscalPolicy:RevenuePolicy:TaxRevenuePolicy:OtherTaxCategories} (Yellow Rows)}}
				}
		};
\end{tikzpicture}
\end{center}
\caption{Atomic policies in \emph{Internation Trade Policy}}
\label{diagram:InternationalTradePolicy}
\end{figure}
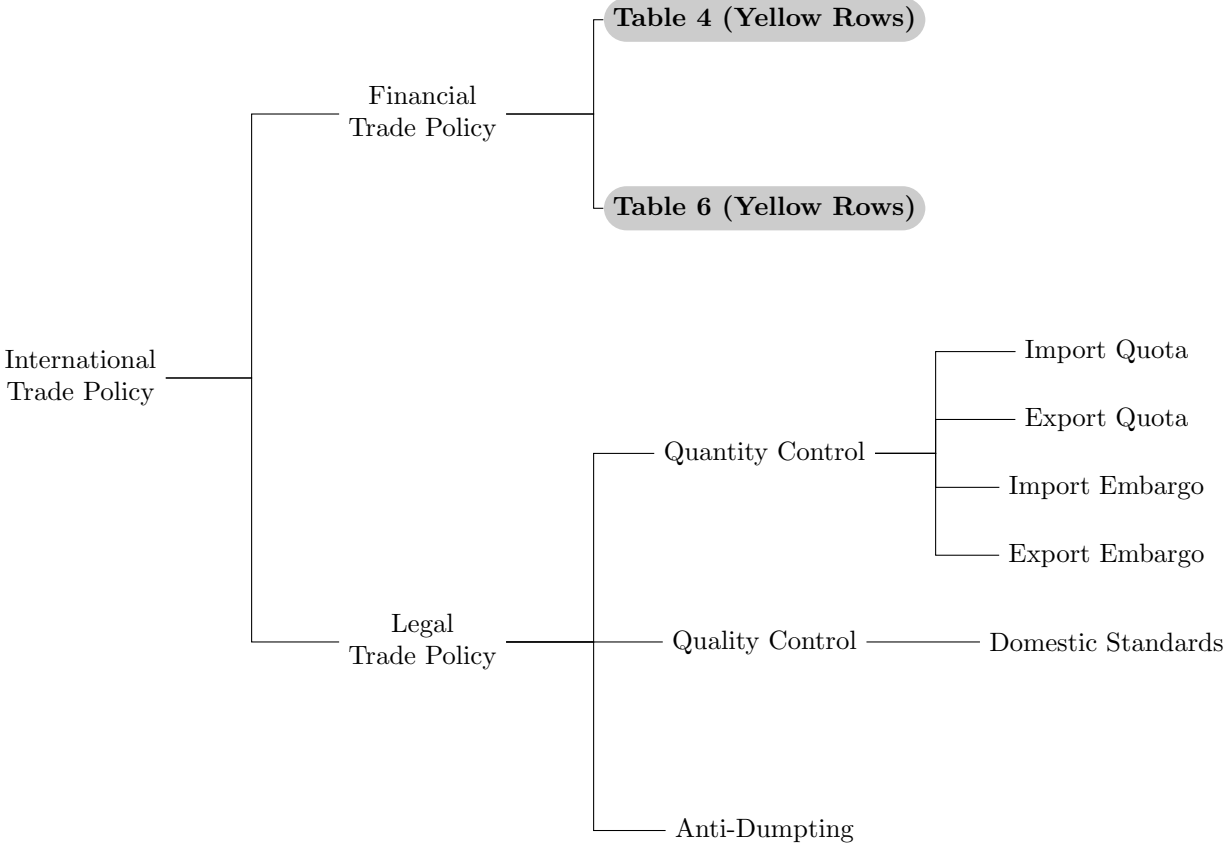

In this paper we introduced a concept of an atomic policy, which is the smallest possible ``interaction'' that a policy maker can perform, and a trait, which is a property that can help us define atomic policies in a convenient way. Using traits we have constucted tables listing the possible atomic policies that a policy maker has access to, and therefore, built a top-to-bottom taxonomy of atomic economic policies. For the purpose of the presentation, we have also provided the taxonomy as a tree diagram in appendix on figure \ref{diagram:EconomicPolicy} (further ramifications of the diagram are presented on figures \ref{diagram:FiscalPolicy}, \ref{diagram:FiscalPolicy:RevenueOperationsPolicy}, \ref{diagram:MonetaryPolicy} and \ref{diagram:InternationalTradePolicy}).

\newpage
\appendix
\appendixpage
\section{Taxonomy Root}
\label{appendix:TaxonomyRoot}
\begin{figure}[h!]
\begin{center}
\footnotesize
\begin{tikzpicture}[
	level 1/.style={sibling distance=60mm},
	level 2/.style={sibling distance=60mm},
	level 3/.style={sibling distance=60mm},
	edge from parent fork down,
]
	\node {Economic Policy}
		child {
			node {\begin{tabular}{@{}c@{}}International \\ Trade Policy\end{tabular}}
				child {node[rounded corners=0.3cm, fill=gray!40] {\textbf{Figure \ref{diagram:InternationalTradePolicy}}}}				
		}
		child {
			node {Stabilization Policy}
				child {
					node {Fiscal Policy}
						child {node[rounded corners=0.3cm, fill=gray!40] {\textbf{Figure \ref{diagram:FiscalPolicy}}}}
				}
				child {
					node {Monetary Policy}
						child {node[rounded corners=0.3cm, fill=gray!40] {\textbf{Figure \ref{diagram:MonetaryPolicy}}}}						
				}
		};
\end{tikzpicture}
\end{center}
\caption{Taxonomy of \emph{Economic Policies}}
\label{diagram:EconomicPolicy}
\end{figure}
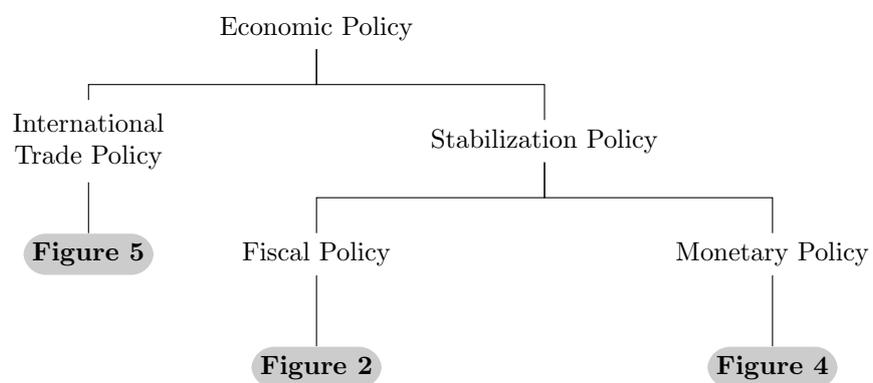

\newpage
\section{Fiscal Policy}
\label{appendix:FiscalPolicy}

\newpage

\newpage
\subsection{Tax Revenue Policy}
\begin{table}[h!]
\centering
\footnotesize
\rotatebox{90}{
\begin{tabular}{ |c|c|c|c|c|c|c|c|c|c| } 
\hline
\textbf{Tax Category} & \begin{tabular}{@{}c@{}c@{}}\textbf{Tax} \\ \textbf{Caluclation} \\ \textbf{Type}\end{tabular} & \begin{tabular}{@{}c@{}}\textbf{Tax} \\ \textbf{Base}\end{tabular} & \begin{tabular}{@{}c@{}c@{}}\textbf{Tax} \\ \textbf{Payment} \\ \textbf{Type}\end{tabular} & \begin{tabular}{@{}c@{}c@{}}\textbf{Tax} \\ \textbf{Evasion} \\ \textbf{Penalty}\end{tabular} & \textbf{Allowance} & \textbf{Abatement} & \textbf{Exemption} & \begin{tabular}{@{}c@{}}\textbf{Negative} \\ \textbf{Tax}\end{tabular} & \begin{tabular}{@{}c@{}}\textbf{Tax} \\ \textbf{Credit}\end{tabular} \\
\hline
\begin{tabular}{@{}c@{}c@{}}Personal \\ Income \\ Tax\end{tabular} & \checkmark &  & \checkmark & \checkmark & \checkmark & \checkmark & \checkmark & \checkmark & \checkmark \\
\hline
\begin{tabular}{@{}c@{}}Capital Gains \\ Tax\end{tabular} & \checkmark &  & \checkmark & \checkmark & \checkmark & \checkmark & \checkmark & \checkmark & \checkmark \\
\hline
\begin{tabular}{@{}c@{}}Dividend \\ Tax\end{tabular} & \checkmark &  & \checkmark & \checkmark & \checkmark & \checkmark & \checkmark & \checkmark & \checkmark \\
\hline
\begin{tabular}{@{}c@{}}Corporate \\ Tax\end{tabular} & \checkmark &  & \checkmark & \checkmark & \checkmark & \checkmark & \checkmark & \checkmark & \checkmark \\
\hline
\begin{tabular}{@{}c@{}c@{}}Excess \\ Profit Tax \\ {[}excess amount{]}\end{tabular} & \checkmark &  & \checkmark & \checkmark &  & \checkmark & \checkmark &  & \\
\hline
\end{tabular}
}
\caption{\emph{Income Tax} categories and their traits}
\label{table:FiscalPolicy:RevenuePolicy:TaxRevenuePolicy:IncomeTax}
\end{table}

\newpage
\begin{table}[h!]
\centering
\footnotesize
\rotatebox{90}{
\begin{tabular}{ |c|c|c|c|c|c|c|c|c|c| } 
\hline
\textbf{Tax Category} & \begin{tabular}{@{}c@{}c@{}}\textbf{Tax} \\ \textbf{Caluclation} \\ \textbf{Type}\end{tabular} & \begin{tabular}{@{}c@{}}\textbf{Tax} \\ \textbf{Base}\end{tabular} & \begin{tabular}{@{}c@{}c@{}}\textbf{Tax} \\ \textbf{Payment} \\ \textbf{Type}\end{tabular} & \begin{tabular}{@{}c@{}c@{}}\textbf{Tax} \\ \textbf{Evasion} \\ \textbf{Penalty}\end{tabular} & \textbf{Allowance} & \textbf{Abatement} & \textbf{Exemption} & \begin{tabular}{@{}c@{}}\textbf{Negative} \\ \textbf{Tax}\end{tabular} & \begin{tabular}{@{}c@{}}\textbf{Tax} \\ \textbf{Credit}\end{tabular} \\
\hline
\begin{tabular}{@{}c@{}}Inheritance \\ Tax\end{tabular} & \checkmark & \checkmark & \checkmark & \checkmark & \checkmark & \checkmark & \checkmark &  & \checkmark \\
\hline
\begin{tabular}{@{}c@{}}Estate \\ Tax\end{tabular} & \checkmark & \checkmark & \checkmark & \checkmark & \checkmark & \checkmark & \checkmark &  & \checkmark \\
\hline
\begin{tabular}{@{}c@{}}Gift \\ Tax\end{tabular} & \checkmark & \checkmark & \checkmark & \checkmark & \checkmark & \checkmark & \checkmark &  & \checkmark \\
\hline
\begin{tabular}{@{}c@{}}Wealth \\ Tax\end{tabular} & \checkmark &  & \checkmark & \checkmark & \checkmark & \checkmark & \checkmark &  & \checkmark \\
\hline
\begin{tabular}{@{}c@{}}Net Wealth \\ Tax\end{tabular} & \checkmark &  & \checkmark & \checkmark & \checkmark & \checkmark & \checkmark &  & \checkmark \\
\hline
\begin{tabular}{@{}c@{}}Real Estate \\ Tax\end{tabular} & \checkmark & \checkmark & \checkmark & \checkmark & \checkmark & \checkmark & \checkmark &  & \checkmark \\
\hline
\begin{tabular}{@{}c@{}}Land Value\\ Tax\end{tabular} & \checkmark & \checkmark & \checkmark & \checkmark & \checkmark & \checkmark & \checkmark &  & \checkmark \\
\hline
\begin{tabular}{@{}c@{}}Expatriation\\ Tax\end{tabular} & \checkmark &  & \checkmark & \checkmark & \checkmark & \checkmark & \checkmark & \checkmark & \checkmark \\
\hline
\begin{tabular}{@{}c@{}}Franchise \\ Tax\end{tabular} & \checkmark &  & \checkmark & \checkmark & \checkmark & \checkmark & \checkmark &  & \checkmark \\
\hline
Carucage & \checkmark &  &  & \checkmark &  &  & \checkmark &  & \checkmark \\
\hline
\end{tabular}
}
\caption{\emph{Property Tax} categories and their traits}
\label{table:FiscalPolicy:RevenuePolicy:TaxRevenuePolicy:PropertyTax}
\end{table}

\newpage
\begin{table}[h!]
\centering
\rotatebox{90}{
\begin{tabular}{ |c|c|c|c|c|c|c|c|c|c| } 
\hline
\textbf{Tax Category} & \begin{tabular}{@{}c@{}c@{}}\textbf{Tax} \\ \textbf{Caluclation} \\ \textbf{Type}\end{tabular} & \begin{tabular}{@{}c@{}}\textbf{Tax} \\ \textbf{Base}\end{tabular} & \begin{tabular}{@{}c@{}c@{}}\textbf{Tax} \\ \textbf{Payment} \\ \textbf{Type}\end{tabular} & \begin{tabular}{@{}c@{}c@{}}\textbf{Tax} \\ \textbf{Evasion} \\ \textbf{Penalty}\end{tabular} & \textbf{Allowance} & \textbf{Abatement} & \textbf{Exemption} & \begin{tabular}{@{}c@{}}\textbf{Negative} \\ \textbf{Tax}\end{tabular} & \begin{tabular}{@{}c@{}}\textbf{Tax} \\ \textbf{Credit}\end{tabular} \\
\hline
\begin{tabular}{@{}c@{}}Value Added \\ Tax\end{tabular} &  & \checkmark &  & \checkmark &  & \checkmark &  &  & \\
\hline
\begin{tabular}{@{}c@{}}Manufacturers' \\ Sale Tax\end{tabular} &  & \checkmark &  & \checkmark & \checkmark & \checkmark & \checkmark &  & \checkmark \\
\hline
\begin{tabular}{@{}c@{}}Wholesale \\ Sales Tax\end{tabular} &  & \checkmark &  & \checkmark & \checkmark & \checkmark & \checkmark &  & \checkmark \\
\hline
\begin{tabular}{@{}c@{}}Retail Sales \\ Tax\end{tabular} &  & \checkmark &  & \checkmark & \checkmark & \checkmark & \checkmark &  & \checkmark \\
\hline
\begin{tabular}{@{}c@{}}Gross Receipts \\ Tax\end{tabular} &  &  &  & \checkmark & \checkmark & \checkmark & \checkmark &  & \checkmark \\
\hline
Luxury Tax &  & \checkmark &  & \checkmark & \checkmark & \checkmark & \checkmark &  & \\
\hline
Excise Tax &  & \checkmark &  & \checkmark & \checkmark & \checkmark & \checkmark &  & \checkmark \\
\hline
Use Tax &  & \checkmark & \checkmark & \checkmark & \checkmark & \checkmark & \checkmark &  & \checkmark \\
\hline
\begin{tabular}{@{}c@{}c@{}}Securities \\ Turnover \\ Excise Tax\end{tabular} & \checkmark &  & \checkmark & \checkmark & \checkmark & \checkmark & \checkmark &  & \checkmark \\
\hline
Turnover Tax & \checkmark & \checkmark & \checkmark & \checkmark & \checkmark & \checkmark & \checkmark &  & \checkmark \\
\hline
\end{tabular}
}
\caption{\emph{Property Tax} categories and their traits}
\label{table:FiscalPolicy:RevenuePolicy:TaxRevenuePolicy:SalesTax}
\end{table}

\newpage
\begin{table}[h!]
\centering
\rotatebox{90}{
\begin{tabular}{ |c|c|c|c|c|c|c|c|c|c| } 
\hline
\textbf{Tax Category} & \begin{tabular}{@{}c@{}c@{}}\textbf{Tax} \\ \textbf{Caluclation} \\ \textbf{Type}\end{tabular} & \begin{tabular}{@{}c@{}}\textbf{Tax} \\ \textbf{Base}\end{tabular} & \begin{tabular}{@{}c@{}c@{}}\textbf{Tax} \\ \textbf{Payment} \\ \textbf{Type}\end{tabular} & \begin{tabular}{@{}c@{}c@{}}\textbf{Tax} \\ \textbf{Evasion} \\ \textbf{Penalty}\end{tabular} & \textbf{Allowance} & \textbf{Abatement} & \textbf{Exemption} & \begin{tabular}{@{}c@{}}\textbf{Negative} \\ \textbf{Tax}\end{tabular} & \begin{tabular}{@{}c@{}}\textbf{Tax} \\ \textbf{Credit}\end{tabular} \\
\hline
\begin{tabular}{@{}c@{}}Environment \\ Tax\end{tabular} & \checkmark & \checkmark &  & \checkmark & \checkmark & \checkmark & \checkmark &  & \checkmark \\
\hline
\begin{tabular}{@{}c@{}}Congestion \\ Tax\end{tabular} & \checkmark & \checkmark &  & \checkmark & \checkmark & \checkmark & \checkmark &  & \checkmark \\
\hline
\begin{tabular}{@{}c@{}}Pigouvian \\ Good Tax\end{tabular} & \checkmark & \checkmark & \checkmark & \checkmark & \checkmark & \checkmark & \checkmark &  & \checkmark \\
\hline
\begin{tabular}{@{}c@{}}Pigouvian \\ Service Tax\end{tabular} & \checkmark & \checkmark & \checkmark & \checkmark & \checkmark & \checkmark & \checkmark &  & \checkmark \\
\hline
Scutage &  &  &  & \checkmark &  & \checkmark & \checkmark &  & \\
\hline
Bank Tax &  &  &  & \checkmark & \checkmark & \checkmark & \checkmark &  & \checkmark \\
\hline
\rowcolor{yellow} Tariff & \checkmark & \checkmark & \checkmark & \checkmark &  & \checkmark & \checkmark &  & \\
\hline
\end{tabular}
}
\caption{Other tax categories (including categories of \emph{Pigouvian Tax}) and their traits}
\label{table:FiscalPolicy:RevenuePolicy:TaxRevenuePolicy:OtherTaxCategories}
\end{table}

\newpage
\subsection{Non-Tax Revenue Policy}
\begin{table}[h!]
\centering
\begin{tabular}{ |c|c| } 
\hline
\textbf{Government's Goods and Services Category} & \textbf{Price Level} \\
\hline
\begin{tabular}{@{}c@{}}State-Owned \\ Enterprise Revenue\end{tabular} & \checkmark \\
\hline
\begin{tabular}{@{}c@{}}Rents and Royalties from \\ Private Companies\end{tabular} & \checkmark \\
\hline
Fines & \checkmark \\
\hline
\begin{tabular}{@{}c@{}}Permits and Licences \\ Fees\end{tabular} & \checkmark \\
\hline
\end{tabular}
\caption{\emph{Government's Goods and Services} and their traits}
\label{table:FiscalPolicy:RevenuePolicy:NonTaxRevenuePolicy:GovernmentGoodsAndServices}
\end{table}

\subsection{Expenditure Policy}
\begin{table}[h!]
\centering
\begin{tabular}{ |c|c| } 
\hline
\textbf{Other Expense Category} & \textbf{Payment Size} \\
\hline
Government Grant & \checkmark \\
\hline
\begin{tabular}{@{}c@{}}Production \\ Subsidization\end{tabular} & \checkmark \\
\hline
\begin{tabular}{@{}c@{}}Consumption \\ Subsidization\end{tabular} & \checkmark \\
\hline
\begin{tabular}{@{}c@{}}Price \\ Subsidization\end{tabular} & \checkmark \\
\hline
\begin{tabular}{@{}c@{}}Employment \\ Subsidization\end{tabular} & \checkmark \\
\hline
\rowcolor{yellow} \begin{tabular}{@{}c@{}}Import \\ Subsidization\end{tabular} & \checkmark \\
\hline
\rowcolor{yellow} \begin{tabular}{@{}c@{}}Export \\ Subsidization\end{tabular} & \checkmark \\
\hline
\end{tabular}
\caption{\emph{Government's Goods and Services} and their traits}
\label{table:FiscalPolicy:ExpenditurePolicy:OtherExpenses}
\end{table}

\newpage
\section{Monetary Policy}
\label{appendix:MonetaryPolicy}

\newpage
\subsection{Open Market Operations Policy}
\begin{table}[h!]
\centering
\begin{tabular}{ |c|c|c| } 
\hline
\textbf{Open Market Operations Category} & \textbf{Security Purchase} & \textbf{Security Sale} \\
\hline
Government Security & \checkmark & \checkmark \\
\hline
Foreign Currency & \checkmark & \checkmark \\
\hline
Commodity & \checkmark & \checkmark \\
\hline
Repurchase Agreement & \checkmark & \checkmark \\
\hline
Reverse Repurchase Agreement & \checkmark & \checkmark \\
\hline
\end{tabular}
\caption{\emph{Open Market Operations Categories} and their traits}
\label{table:MonetaryPolicy:OpenMarketOperationsPolicy}
\end{table}

\subsection{Debt and Credit Policy}
\begin{table}[h!]
\centering
\begin{tabular}{ |c|c|c|c|c|c|c| } 
\hline
\begin{tabular}{@{}c@{}}\textbf{Debtors/} \\ \textbf{Creditors}\end{tabular} & \begin{tabular}{@{}c@{}}\textbf{Credit} \\ \textbf{Issuance}\end{tabular} & \begin{tabular}{@{}c@{}}\textbf{Credit} \\ \textbf{Quota}\end{tabular} & \begin{tabular}{@{}c@{}}\textbf{Credit} \\ \textbf{Embargo}\end{tabular} & \begin{tabular}{@{}c@{}c@{}}\textbf{Differential} \\ \textbf{Interest} \\ \textbf{Rate}\end{tabular} & \begin{tabular}{@{}c@{}}\textbf{Debt} \\ \textbf{Issuance}\end{tabular} & \begin{tabular}{@{}c@{}}\textbf{Debt} \\ \textbf{Default}\end{tabular} \\
\hline
Sovereign & \checkmark & \checkmark& \checkmark & \checkmark & \checkmark & \checkmark \\
\hline
\begin{tabular}{@{}c@{}}Foreign \\ Business\end{tabular} &  &  &  &  & \checkmark & \checkmark \\
\hline
\begin{tabular}{@{}c@{}}Foreign \\ Household\end{tabular} &  &  &  &  & \checkmark & \checkmark \\
\hline
\begin{tabular}{@{}c@{}}Domestic \\ Business\end{tabular} & \checkmark & \checkmark & \checkmark & \checkmark & \checkmark & \checkmark \\
\hline
\begin{tabular}{@{}c@{}}Domestic \\ Household\end{tabular} & \checkmark & \checkmark & \checkmark & \checkmark & \checkmark &\checkmark \\
\hline
\end{tabular}
\caption{\emph{Debtor/Creditor Categories} and their traits}
\label{table:MonetaryPolicy:DebtAndCreditPolicy}
\end{table}

\subsection{Financial Markets Policy}
\begin{table}[h!]
\centering
\begin{tabular}{ |c|c|c|c|c| } 
\hline
\textbf{Financial Markets} & \begin{tabular}{@{}c@{}}\textbf{Administered} \\ \textbf{Rate}\end{tabular} & \begin{tabular}{@{}c@{}}\textbf{Rate} \\ \textbf{Corridor}\end{tabular} & \begin{tabular}{@{}c@{}}\textbf{Collateral} \\ \textbf{Assets}\end{tabular} & \begin{tabular}{@{}c@{}}\textbf{Margin} \\ \textbf{Requirement}\end{tabular} \\
\hline
\begin{tabular}{@{}c@{}}Interbank \\ Refinancing\end{tabular} & \checkmark & \checkmark &  & \\
\hline
\begin{tabular}{@{}c@{}}Repurchase \\ Agreement\end{tabular} & \checkmark & \checkmark & \checkmark & \checkmark \\
\hline
\end{tabular}
\caption{\emph{Financial Markets} and their traits}
\label{table:MonetaryPolicy:FinancialMarketsPolicy}
\end{table}

\newpage
\section{International Trade Policy}
\label{appendix:InternationalTradePolicy}

\newpage
\section{Additional Features}
\label{appendix:AdditionalFeatures}

\newpage
\begin{thebibliography}{99}
	
	\bibitem{IFRSConceptualFramework} IFRS Foundation, 2025. \emph{IFRS Conceptual Framework}, 2018. \url{https://www.ifrs.org/issued-standards/list-of-standards/conceptual-framework.html/content/dam/ifrs/publications/html-standards/english/2025/issued/cf/#about}
	
	\bibitem{MacroMicroInterlockedSimulator} T. Sato, 2005. \emph{Macro-Micro Interlocked Simulator}, Journal of Physics: Conference Series, Vol. 16, page 310, June 2005. \url{https://iopscience.iop.org/article/10.1088/1742-6596/16/1/043}
	
	\bibitem{MacroMicroEconomicSystemSimulation} B.S. Onggo, K. Kusano and T. Sato, 2007. \emph{Macro-Micro Economic System Simulation}, Proceedings-Workshop on Principles of Advanced and Distributed Simulation (PADS'07), pages 105-112, July 2007. \url{https://www.doi.org/10.1109/PADS.2007.22}
	
	\bibitem{CambridgeDictionary} Cambridge University, 2025. \emph{Cambridge Dictionary}, Cambridge University, Press \& Assessment, 2025. \url{https://dictionary.cambridge.org/dictionary/}
	
	
\end{thebibliography}
\end{document}